\pdfoutput=1
\newcommand*{\ATLASLATEXPATH}{}
\documentclass[PAPER, UKenglish, cernpreprint, texlive=2016]{\ATLASLATEXPATH atlasdoc}
 
\usepackage{\ATLASLATEXPATH atlaspackage}
\usepackage{\ATLASLATEXPATH atlasbiblatex}
 
\usepackage{\ATLASLATEXPATH atlasphysics}
 
\graphicspath{{logos/}{figures/}}
 
\usepackage{ANA-EXOT-2017-16-PAPER-defs}
 
\usepackage{multirow}

\AtlasTitle{Search for large missing transverse momentum in association with one top-quark in proton--proton collisions at $\sqrt{s}=13$~\TeV\ with the ATLAS detector}
 
\AtlasAbstract{
This paper describes a search for events with one top-quark and large missing transverse momentum in
the final state.  Data collected during 2015 and 2016 by the ATLAS experiment from 13~\TeV\
proton--proton collisions at the LHC corresponding to an integrated
luminosity of 36.1~fb$^{-1}$ are used. Two channels are considered, depending on the leptonic or the hadronic decays of the $W$ boson from the top quark. The obtained results are interpreted in the context of
simplified models for dark-matter production and for the single production
of a vector-like $T$ quark. In the absence of significant deviations from the Standard Model
background expectation, 95\% confidence-level upper limits on the corresponding
production cross-sections are obtained and these limits are translated into
constraints on the parameter space of the models considered.
}
 
\author{The ATLAS Collaboration}
 
\AtlasRefCode{EXOT-2017-16}
 
\PreprintIdNumber{CERN-EP-2018-301}

\AtlasJournal{JHEP}
\AtlasJournalRef{JHEP 05 (2019) 41}
\AtlasDOI{10.1007/JHEP05(2019)041}
\hypersetup{pdftitle={ATLAS document},pdfauthor={The ATLAS Collaboration}}
 
\begin{document}
 
\maketitle
 
\tableofcontents

\section{Introduction}
\label{sec:intro}
In spite of its successes in describing the phenomenology of the fundamental particles
and the corresponding interactions, the Standard Model (SM) can be considered
as a low-energy approximation of a more fundamental theory with
new degrees of freedom and symmetries that would become manifest at a higher energy.
 
One argument supporting the idea that new particles beyond the SM might exist arises from
astrophysical measurements, such as the rotational speed of stars in
galaxies and gravitational lensing~\cite{Trimble:1987ee, Bertone:2004pz,
Feng:2010gw}. These observations point to the existence of non-light-emitting
matter, a dominant fraction of which is of non-baryonic form, usually referred
to as dark matter (DM).  Even if there are no viable candidates in the SM for
particles which could explain DM, proton--proton collisions
at the Large Hadron Collider (LHC) may possibly produce new particles that
couple both to SM particles and to these DM candidates. While such candidates are
not expected to interact significantly with detectors, the SM particles produced in
association with the unobserved DM particles could allow these processes to be detected.
Search strategies depend on the type of particle or system that is
recoiling against the unseen particle. Both ATLAS and CMS have carried out
searches for invisible particles produced in association with
jets~\cite{Aaboud:2017phn, Sirunyan:2017hci, Aaboud:2017rzf, Aaboud:2017aeu},
photons~\cite{Aaboud:2017dor, Sirunyan:2017ewk}, $W$ or $Z$ bosons~\cite{Aaboud:2018xdl,
Sirunyan:2017hci, Aaboud:2017bja} and Higgs bosons~\cite{Aaboud:2017yqz, Aaboud:2017uak,
Sirunyan:2017hnk, Sirunyan:2018fpy}, significantly constraining the allowed parameter space
for different classes of models predicting DM candidates.
 
This paper describes a search for the production of invisible particles in
association with a single top-quark in proton--proton collisions produced at
the LHC with a centre-of-mass energy of $\sqrt{s}=13$~\TeV\ and detected using the
ATLAS detector. Such a final state, commonly referred to as ``mono-top'', is
characterised by a top-quark and significant missing transverse momentum,
which is due to the undetected particles. Background contributions from
SM processes~\cite{Andrea:2011ws} are expected to be small. In addition, this search is
sensitive to specific DM models, since the presence of top-quarks in the final
state constrains the flavour structure of the considered
couplings~\cite{Lin:2013sca,Cacciapaglia:2018rqf}. Similar searches were previously conducted by the CDF
Collaboration using 7.7~fb$^{-1}$ of Tevatron $p\bar{p}$ collisions at
$\sqrt{s}=1.96$~\TeV~\cite{Aaltonen:2012ek} and by the ATLAS and CMS collaborations
using $\sqrt{s}=8$~\TeV~\cite{Khachatryan:2014uma,Aad:2014wza} and
$\sqrt{s}=13$~\TeV~\cite{Sirunyan:2018gka} LHC data. Searches for new phenomena in
events with same-charge leptons and $b$-tagged jets~\cite{Aaboud:2018xpj} provide
information complementary to the results from mono-top searches and exclude new
vector resonances with masses up to 3~\TeV, assuming a dark-sector coupling of 1.0 and
a coupling to SM particles above 0.3.
 
A final state with a top-quark and missing transverse momentum can also
originate from the single production of new vector-like quarks if these decay
into a top-quark and a $Z$ boson that decays invisibly into two
neutrinos. Vector-like quarks are colour-triplet spin-$1/2$ fermions in which, in
contrast to the SM quarks, the left- and right-handed components have the same
properties under transformations of the electroweak symmetry group
$\mathrm{SU}(2)_\mathrm{L} \times \mathrm{U}(1)_\mathrm{Y}$.
Such new particles are predicted in SM extensions, such as
Little Higgs~\cite{LittleHiggs,LittleHiggsRev} and Composite
Higgs~\cite{CompHiggs1,CompHiggs2} models, and are expected to mix with SM
quarks~\cite{VLQmixing}. In order to preserve gauge invariance, only a limited set
of possible representations exist~\cite{delAguila,delAguila:2000rc} and their
electric charge can be $+2/3e$ ($T$ quark), $-1/3e$ ($B$ quark), $+5/3e$ ($X$
quark) or $-4/3e$ ($Y$ quark), with $e$ being the elementary charge. In this
paper, only the single production of vector-like $T$ quarks (VLT) via an
electroweak interaction is considered. Although couplings of $T$ quarks to
first- and second-generation SM quarks are not
excluded~\cite{Atre:2008iu,Atre:2011ae}, it is common to assume that they couple
exclusively to third-generation SM quarks~\cite{AguilarSaavedra:2009es}. Such
couplings can be described in terms of
$\sin\theta_\mathrm{L}$~\cite{Aguilar-Saavedra:2013qpa}, where $\theta_\mathrm{L}$
is the mixing angle of the $T$ quark with the top-quark, or in terms of a generalised coupling
$\kappa_T$~\cite{Buchkremer:2013bha,Matsedonskyi:2014mna}. The $T$ quarks
can decay either via the charged current, i.e.\ $T\rightarrow Wb$, or via
flavour-changing neutral currents~\cite{delAguila2}, i.e.\ $T\rightarrow Zt$ and
$T\rightarrow Ht$. The $T\rightarrow Zt \to \nu\bar \nu W b$ decay is considered
in the present search.
 
The ATLAS and CMS collaborations have sought pair production of $T$ quarks
decaying into third-generation quarks in $pp$ collisions at a centre-of-mass energy
of 8~\TeV~\cite{Aad:2015kqa,Aad:2015gdg,Aad:2014efa,Khachatryan:2015oba},
targeting all three possible decay modes. Searches at 13~\TeV\ have
aimed at final states with leptons, targeting the $T\rightarrow Zt$
decay~\cite{Aaboud:2017qpr,Aaboud:2018saj}, the $T\rightarrow
Wb$ decay~\cite{Aaboud:2017zfn,Sirunyan:2017pks}, as well as general single-lepton and
fully hadronic final states with boosted
bosons~\cite{Sirunyan:2017usq,PhysRevD.98.092005} and multiple $b$-tagged
jets~\cite{Sirunyan:2017usq,Aaboud:2018xuw,Sirunyan:2018omb}. The most stringent
mass limit for an isospin singlet $T$ is 1.3~\TeV~\cite{PhysRevLett.121.211801}. For such
large $T$ masses, the cross-section for single $T$ production may be larger
than the pair-production cross-section because of the larger available phase
space. Nonetheless, the comparison of single- and pair-production cross-sections
depends on the assumed coupling to the SM quarks. Single production of $T$ quarks
was sought at 8~\TeV~\cite{Aad:2016qpo,Aad:2015voa,Aad:2014efa} by the ATLAS
Collaboration. At 13~\TeV, the ATLAS and CMS collaborations have sought the
decays $T\rightarrow Wb$~\cite{Sirunyan:2017tfc,Sirunyan:2017bfa}, $T\rightarrow
Ht$~\cite{Sirunyan:2016ipo,Khachatryan:2016vph} and $T\rightarrow
Zt$~\cite{Sirunyan:2017ynj,Sirunyan:2017ezy,Aaboud:2018saj}.
 
In this paper, two channels for the mono-top signature are considered,
targeting the case in which the \Wboson~boson originating from the top-quark
decays into an electron or muon and a neutrino (leptonic channel) and the case in which it
decays into a pair of quarks (hadronic channel). These analyses define different
signal regions, maximising the signal discovery sensitivity, and control regions,
enriched with the dominant background processes. The statistical interpretation of
the results is based on a simultaneous fit to the signal and control regions to
determine a possible signal contribution and constrain the main backgrounds with
data, taking into account experimental and theoretical systematic uncertainties.
 
The paper is organised as follows. The signal models are introduced in Section~\ref{sec:signalPheno}.
After a brief introduction to the ATLAS detector, given in Section~\ref{sec:detector}, the data samples
and samples of simulated signal and background events are described in Section~\ref{sec:data}.
The algorithms for the reconstruction and identification of final-state particles are summarised in Section~\ref{sec:objects}.
Section~\ref{sec:selection} describes the criteria for the selection of candidate signal events.
This section also describes the estimation of the background contribution with the help of dedicated control regions in data.
The experimental and theoretical systematic uncertainties (Section~\ref{sec:systematic}) are taken
into account in the statistical interpretation of data, with the results presented in Section~\ref{sec:results}.
Concluding remarks are given in Section~\ref{sec:conclusions}.

\section{Signal phenomenology} \label{sec:signalPheno}
 
This paper presents a search for two different signals: DM candidates
produced in association with top-quarks and single production of vector-like $T$
quarks decaying into a top-quark and a $Z$ boson.
 
\subsection{DM candidates associated with top-quarks}
 
\begin{figure}[!htp]
\centering
\begin{tabular}{c c}
\includegraphics[width=0.42\textwidth]{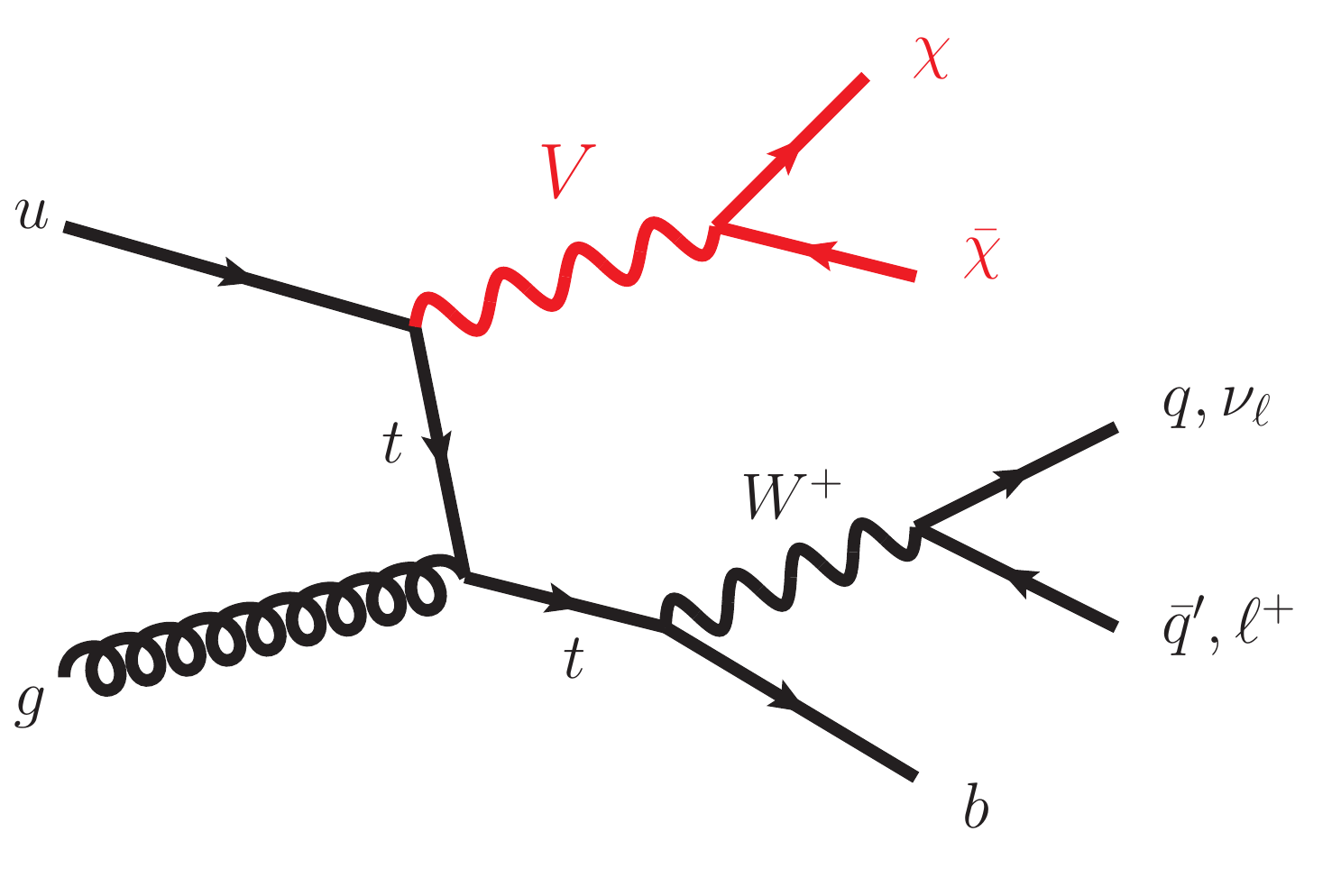} &
\includegraphics[width=0.42\textwidth]{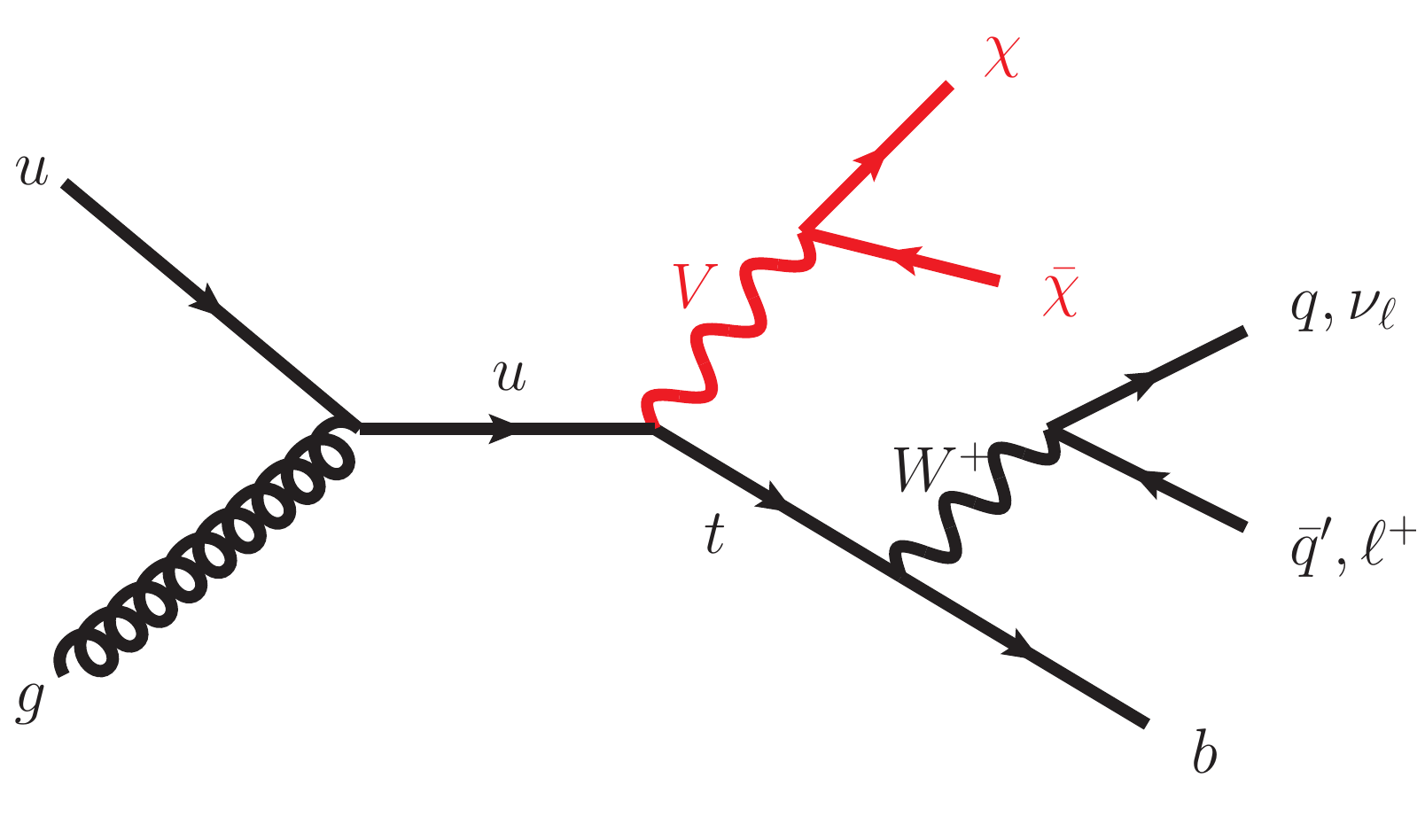} \\[-2mm]
(a) & (b) \\[2mm]
\includegraphics[width=0.42\textwidth]{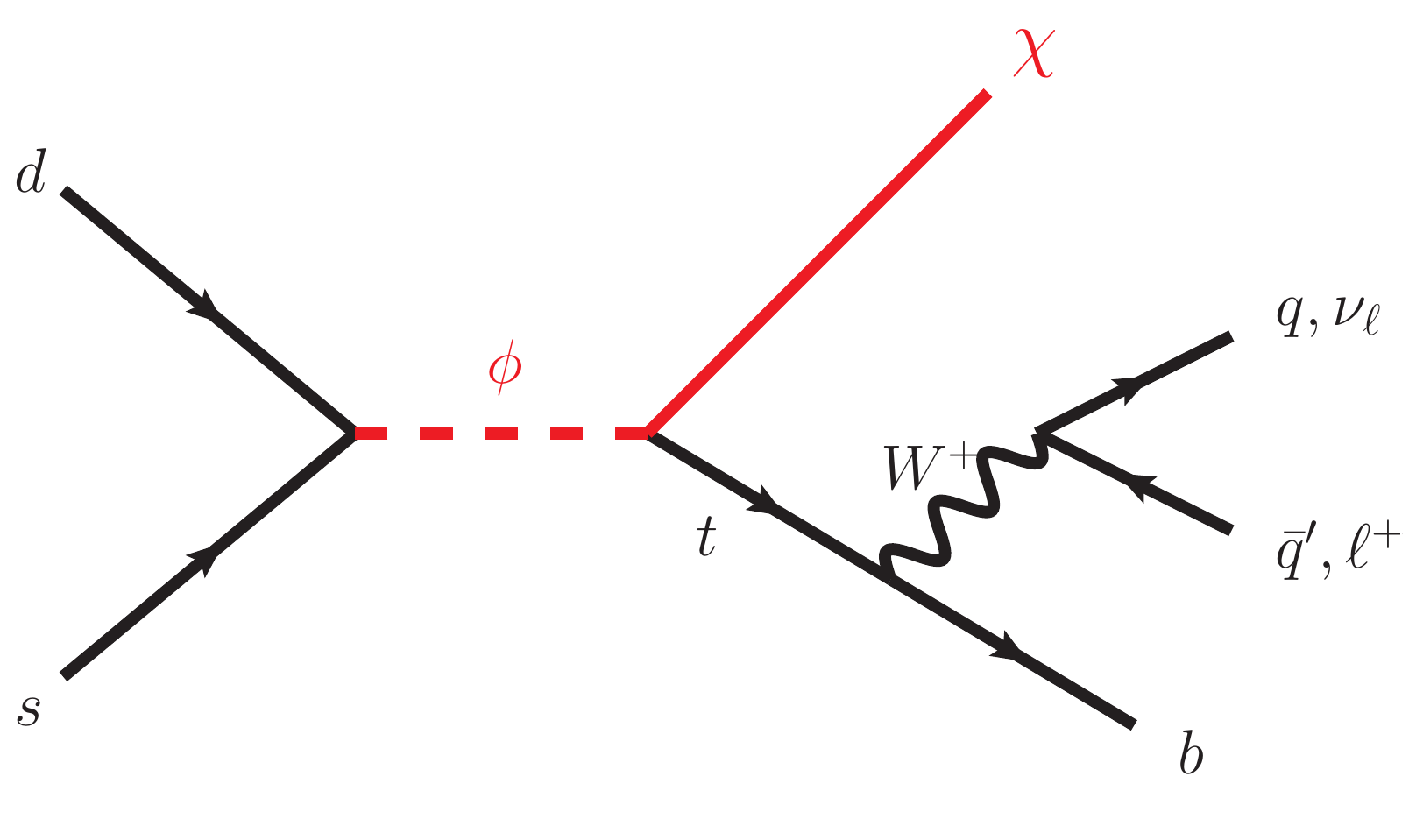} &
\includegraphics[width=0.42\textwidth]{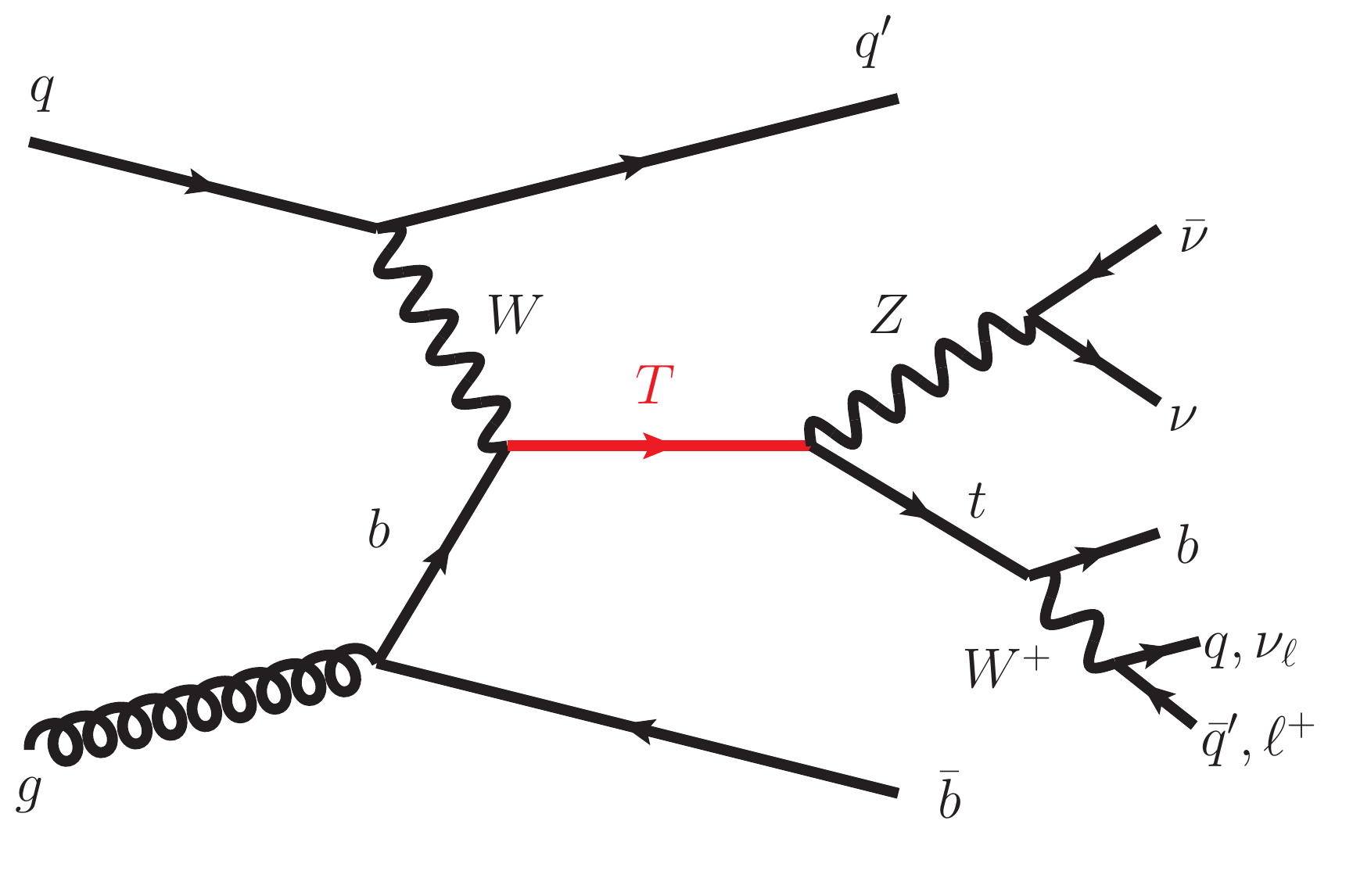} \\[-2mm]
(c) & (d) \\
\end{tabular}
\caption{Representative leading-order diagrams corresponding to the signals sought in this paper: non-resonant (a) $t$-channel and (b) $s$-channel production of a top-quark in association with a vector
boson~\textit{V} which decays into two DM particles; (c) resonant production of a coloured scalar~$\phi$ that decays into a DM particle and a top-quark; and (d) single production of a vector-like $T$ quark
decaying into $Zt\, (\to \nu\bar \nu b W)$.}
\label{fig:FeynmanDiagrams}
\end{figure}

In this search the resonant and non-resonant production of DM
particles associated with a top-quark are considered. The non-resonant
case, represented in \Fig{\ref{fig:FeynmanDiagrams}(a)} and \Fig{\ref{fig:FeynmanDiagrams}(b)},
corresponds to a flavour-changing neutral-current
interaction, producing a top-quark and a new vector particle $V$, which in turn
decays invisibly into a pair of DM particles. Such a process can be parameterised
through a general Lagrangian~\cite{Abercrombie:2015wmb,Andrea:2011ws}:
\begin{equation*}
\mathcal{L}_{\mathrm{int}} =  a V_{\mu} \bar{u}\gamma^{\mu} P_{\mathrm{R}} t + g_{\chi} V_{\mu} \bar{\chi} \gamma^{\mu}\chi + \mathrm{h.c.}\,,
\end{equation*}
where a massive vector boson $V$ is coupled to a DM particle (represented by a Dirac fermion $\chi$)
with a strength controlled by the parameter $g_{\chi}$.
The term $P_{\mathrm{R}}$ is the right-handed chirality projector.
The parameter $a$ stands for the coupling constant between the massive vector boson $V$ and
the $t$- and $u$-quarks, and $\gamma^\mu$ are the Dirac matrices.
Another possibility is the resonant case, corresponding to the
production of a coloured charge-2/3 scalar ($\phi$) decaying into a top-quark
and a spin-1/2 DM particle ($\chi$)~\cite{Wang:2011uxa}. This process, represented
in \Fig{\ref{fig:FeynmanDiagrams}(c)}, is described by the following
Lagrangian~\cite{Abercrombie:2015wmb,Andrea:2011ws}:
\begin{equation*}
\mathcal{L}_{\mathrm{int}} =  \lambda\phi \bar{d}^{c} P_{\mathrm{R}} s + y\phi\bar{\chi} P_{\mathrm{R}} t + \mathrm{h.c.}\,,
\end{equation*}
where the parameters $\lambda$ and $y$ represent the couplings of the charged
scalar to the $d$- and $s$-quarks and to the top-quark and the DM particle $\chi$,
respectively.

\subsection{Single production of vector-like $T$ quarks}
 
The single production of $T$ quarks can occur via a charged $WTb$ or a neutral
$ZTt$ vertex. However, $ZTt$ production is suppressed because of the required top
quark in the initial state. For this reason, $ZTt$ production is not
considered in this analysis and single VLT production refers to $T$
production via the $WTb$ vertex throughout this paper. The $T$ quarks can decay into
$bW$, $tH$ and $tZ$, with the corresponding branching ratios ($\mathcal{B}$) depending on the
specific model considered~\cite{AguilarSaavedra:2009es,Matsedonskyi:2014mna}.
 
The specific case of single production of vector-like $T$ quarks decaying
into $tZ$, followed by the $Z$ boson decaying into neutrinos, results in a mono-top signature.
As can be seen in \Fig{\ref{fig:FeynmanDiagrams}(d)}, one important
difference between DM production and vector-like $T$-quark production is the presence of
additional quarks in the single production of $T$ quarks, which will lead to at
least one jet being detected at a small angle relative to the beam line.
Similarly to the DM case, the topology of the VLT signal has a
distinctive signature, characterised by the presence of a top-quark and missing
transverse momentum, arising from the $Z \to \nu\bar\nu$ decay (and from the
$t\to bW \to \ell\nu$ decay in the single-lepton channel case).


\section{ATLAS detector}
\label{sec:detector}
 
\newcommand{\AtlasCoordFootnote}{
ATLAS uses a right-handed coordinate system with its origin at the nominal interaction point (IP)
in the centre of the detector and the $z$-axis along the beam pipe.
The $x$-axis points from the IP to the centre of the LHC ring,
and the $y$-axis points upwards.
Cylindrical coordinates $(r,\phi)$ are used in the transverse plane,
$\phi$ being the azimuthal angle around the $z$-axis.
The pseudorapidity is defined in terms of the polar angle $\theta$ as $\eta = -\ln \tan(\theta/2)$.
Angular distance is measured in units of $\Delta R \equiv
\sqrt{(\Delta\eta)^{2} + (\Delta\phi)^{2}}$.}

The ATLAS experiment~\cite{PERF-2007-01} at the LHC is a multipurpose
particle detector with nearly $4\pi$ coverage around the collision
point.\footnote{\AtlasCoordFootnote} It consists of an inner tracking
detector surrounded by a thin superconducting solenoid providing a
\SI{2}{\tesla} axial magnetic field, electromagnetic and hadron
calorimeters, and a muon spectrometer.  The inner tracking detector
covers the pseudorapidity range $|\eta| < 2.5$.  It consists of
a silicon pixel detector, including the insertable B-layer~\cite{ATLAS-TDR-19,1748-0221-13-05-T05008} installed
after Run~1 of the LHC, a silicon microstrip
detector, and a transition-radiation tracking detector.
Lead/liquid-argon (LAr) sampling calorimeters provide electromagnetic
(EM) energy measurements with high granularity for $|\eta| < 3.2$.
A steel/scintillator-tile hadron calorimeter covers the central
pseudorapidity range ($|\eta| < 1.7$).  The endcap and forward
regions are instrumented with LAr calorimeters for both the EM and
hadronic energy measurements up to $|\eta| = 4.9$. The outer part of
the detector includes a muon spectrometer with high-precision tracking
chambers providing coverage up to $|\eta| = 2.7$, fast detectors for
triggering over $|\eta| < 2.4$, and three large air-core toroidal
superconducting magnets with eight coils each. A two-level trigger
system~\cite{TRIG-2016-01}, using custom hardware followed by a
software-based trigger level, is used to select events of interest at an
average rate of \SI{1}{kHz}.

 
\section{Data and simulation samples}
\label{sec:data}
 
This analysis is performed using \pp collision data recorded at a
centre-of-mass energy of $\sqs=\SI{13}{\TeV}$ with the ATLAS detector
during 2015 and 2016 in the periods when the LHC was operating with
\SI{25}{ns} bunch spacing and with an average number of collisions per
bunch crossing $\langle\mu\rangle$ of around 23. Only periods in which
all detector components necessary for this analysis were functional are considered, resulting in a data
sample with a total integrated luminosity of \lumi{}.
 
In the single-lepton channel, events are required to pass at least one
of the single-muon or single-electron
triggers~\cite{TRIG-2016-01}. The triggers require a \pt{} of at least
\SI{20}{\GeV} (\SI{26}{\GeV}) for muons and \SI{24}{\GeV}
(\SI{26}{\GeV}) for electrons for the 2015 (2016) data sets, and also
have requirements on lepton reconstruction and isolation. These are
complemented by triggers with higher \pt{} thresholds and relaxed
isolation and identification requirements to ensure maximum efficiency
at higher lepton \pt. In the hadronic channel, events are considered if
they are accepted by triggers that select events with high missing
transverse momentum, with online thresholds of \SI{70}{\GeV} in 2015
and \SI{90}{\GeV} to \SI{110}{\GeV} in 2016.
 
For all signal and background processes of interest, Monte Carlo (MC) events were simulated.
 
Signal events for both the resonant and non-resonant DM scenarios were generated
according to a simplified model~\cite{Boucheneb:2014wza} described in
Section~\ref{sec:intro}, implemented in \MGMCatNLO{v2.3.2}~\cite{Alwall:2014hca} through
{\texttt FeynRules~2.0}~\cite{Alloul:2013bka, Degrande:2011ua}. Such generation
was done at leading order (LO) using the NNPDF3.0LO~\cite{Ball:2014uwa} parton distribution
function (PDF) set. Parton showering, hadronisation and underlying-event modelling
were handled using the \PYTHIAV{8.212}~\cite{Sjostrand:2007gs} event generator
with the A14~\cite{ATL-PHYS-PUB-2014-021} set of tuned parameters, using the
NNPDF2.3LO PDF set~\cite{Ball:2012cx}. Signal samples for the resonant model were
generated assuming a DM mass of $m_\chi = 10$~\GeV\ and a range of the new scalar
masses, $m_{\phi}$, between $1$~\TeV\ and $5$~\TeV, representing two different kinematic regimes.
The kinematic distributions predicted by the model have only a small dependence on the coupling
parameters and therefore all samples were generated using a coupling constant of
$\lambda = 0.2$ and a mixing parameter of $y=0.4$.
The remaining kinematic dependence on the different couplings and masses was
accounted for by means of a reweighting procedure (see \Sect{\ref{sec:results}} for details).
Similarly, the signal samples for the non-resonant model were generated for values of
$m_{V}$ between $500$~\GeV\ and $3$~\TeV, corresponding to the expected sensitivity of
the analysis, and a benchmark DM mass $m_{\chi}=1$~\GeV. The values of the couplings
were set to $a = 0.5$ and $g_\chi=1.0$. The kinematic effect of changing the
coupling values was taken into account by using the previously mentioned reweighting
procedure. The samples were normalised to the theoretical LO cross-sections,
computed with \MGMCatNLO.

The single production of $T$ quarks was generated using the {\texttt Feynrules 2.0}
implementation of a general model~\cite{Buchkremer:2013bha} interfaced to
\MGMCatNLO{v2.3.2}. The NNPDF2.3 LO PDF set and \PYTHIAV{8.212} with the A14 set of
tuned parameters were used. Since the current analysis targets a final state with
large missing transverse momentum, only the $T\rightarrow Zt$ decay, with $Z$
decaying invisibly, was considered, as represented in
\Fig{\ref{fig:FeynmanDiagrams}(d)}. Samples were generated for $T$ masses in the
range from 700 to 2000~\GeV\ with a benchmark coupling of $\kappa_T = 0.5$ in the
$WTb$ production vertex. Additional samples were generated with alternative values
of $\kappa_T = 0.1$ and $1.0$ in order to study the effect of a varying $T$-quark
width on kinematic distributions. The samples were normalised to the next-to-leading-order (NLO)
cross-section by correcting the LO cross-sections calculated with \MGMCatNLO for the
difference between the NLO and LO cross-sections reported for the neutral single-$T$
production process via a $ZTt$ coupling~\cite{Matsedonskyi:2014mna}. For large
values of the coupling the narrow-width approximation used in the cross-section
calculation does not apply, so the cross-sections were corrected to include
width effects, using a reweighting procedure similar to that previously
mentioned, in order to account for the corresponding kinematic effects.
 
For the background samples, several matrix element (ME) event generators
were combined with parton shower and hadronisation programs.
\POWHEGBOX{ v2}~\cite{Nason:2004rx,Frixione:2007vw,Alioli:2010xd,Frixione:2007nw,Frederix:2012dh,Re:2010bp,Alioli:2009je}
interfaced to \PYTHIAV{8.210} using the A14 set of tuned parameters
was used to simulate \ttbar production at NLO.
Single top production was generated at NLO with \POWHEGBOX{ v1} for the $t$-, $Wt$- and $s$-channels and at LO with \MGMCatNLO for the $tZq$ process, interfaced to \PYTHIAV{6.428}~\cite{pythia6}. The CTEQ6L1 PDF set~\cite{cteq6} and the Perugia 2012 set of tuned parameters~\cite{perugia} were used in the parton shower,
hadronisation, and underlying-event simulation. The CT10f4 (CT10) PDF
set~\cite{ct10} was used in the matrix element calculations for the $t$-channel ($Wt$- and $s$- channels).
To model the \Wjets
and \Zjets background the \SHERPA{ v2.2.1}~\cite{sherpa} generator was
used. Matrix elements were calculated for up to two partons at \NLO and
up to four partons at \LO using the \textsc{Comix}~\cite{comix}
and \textsc{OpenLoops}~\cite{openloops} ME generators, and merged with
the \SHERPA parton shower~\cite{csshower} according to the \textsc{ME+PS@NLO}
prescription~\cite{mepsnlo}. The NNPDF3.0 next-to-NLO (NNLO) PDF set~\cite{nnpdf3.0}
was used in conjunction with a \SHERPA parton shower tuning from the authors.
Diboson processes were simulated with \POWHEGBOX{ v2} interfaced
to \PYTHIAV{8.186}. The CT10nlo PDF set was used for the hard process
while the CTEQ6L1 PDF set was used for the parton shower. For the
simulation of \ttbar events with additional bosons
\ttbar+\,$X$ ($X=\Wboson$, \Zboson, Higgs), \MGMCatNLOV{v2.3.2}
interfaced to \PYTHIAV{8.186} was used at NLO in QCD. Non-perturbative effects were
modelled with the AZNLO set of tuned parameters~\cite{STDM-2012-23}.

The considered cross-sections for the dominant backgrounds, \ttbar and \Wboson/\Zboson+\,jets, were evaluated
at \NNLO in quantum chromodynamics (QCD)~\cite{top++,Catani:2009sm}. The calculation for \ttbar{} also includes next-to-
next-to-leading logarithmic soft gluon terms.
 
The \textsc{EvtGen}~v1.2.0 program~\cite{evtgen} was used
to simulate properties of the bottom and charmed hadron decays except
for samples generated with \SHERPA.  All
simulated samples except the DM non-resonant signal in the
leptonic channel and \ttbar+\,$X$ processes were processed with the full
simulation of the ATLAS detector~\cite{SOFT-2010-01}
using \GEANT{4}~\cite{geant4}.
Additional samples used in the estimation of systematic uncertainties
were instead produced using \textsc{Atlfast2}~\cite{ATL-PHYS-PUB-2010-013}, in which a
parameterised detector simulation was used for the calorimeter
responses. This simulation was also used for the generation of the DM
non-resonant signal in the leptonic channel and \ttbar+\,$X$
processes.
All samples were simulated with a varying number of
minimum-bias interactions generated
with \PYTHIAV{8.186} using the A2 set of tuned parameters~\cite{ATL-PHYS-PUB-2012-003},
overlaid on the hard-scattering
event to account for the multiple \pp interactions in the same or
nearby bunch crossings (pile-up). Simulated events were corrected using per-event weights to describe the distribution of the average number of interactions per proton bunch-crossing as observed in data
 
\section{Event reconstruction and object selection}
\label{sec:objects}
 
Events are required to have at least one vertex candidate with at
least two tracks with $\pt>\SI{400}{\MeV}$. The primary vertex is
taken to be the vertex candidate with the largest sum of squared
transverse momenta of all associated tracks.
 
Electron candidates are reconstructed from an isolated electromagnetic
calorimeter energy deposit matched to a track in the inner detector
passing tight likelihood-based
requirements~\cite{ATLAS-CONF-2016-024}. They are required to have a
transverse energy $\et>\SI{30}{\GeV}$ and pseudorapidity
$|\eta|<2.47$, with the transition region between the barrel and
endcap electromagnetic calorimeters, $1.37<|\eta|<1.52$, excluded.
Electron candidates must have a track satisfying requirements of
$|d_0|/\sigma_{d_{0}} < 5$ for the transverse impact parameter significance
relative to the beamline and $|\Delta z_0 \sin\theta |<\SI{0.5}{mm}$ for the
longitudinal impact parameter calculated relative to the primary vertex.
Furthermore, electrons must satisfy isolation
requirements based on inner detector tracks and topological clusters
in the calorimeter~\cite{PERF-2014-07}, with an isolation efficiency
of $90\%$~($99\%$) for electrons from \Zee decays with
$\pt=\SI[parse-numbers=false]{25 (60)}{\GeV}$. Correction factors are
applied to simulated electrons to take into account the small
differences in reconstruction, identification, and isolation
efficiencies between data and MC simulation.
 
Muon candidates are reconstructed by combining tracks reconstructed in the
inner detector with matching tracks reconstructed in the muon spectrometer, and
are required to satisfy $\pt>\SI{30}{\GeV}$ and $|\eta|<2.5$
~\cite{PERF-2015-10}. Muon candidates must satisfy requirements of
$|d_0|/\sigma_{d_{0}} < 3$ and $|\Delta z_0 \sin \theta |<\SI{0.5}{mm}$
for the transverse impact parameter significance and
the longitudinal impact parameter, respectively.
An isolation requirement based on inner detector tracks and topological
clusters in the calorimeters is imposed, which achieves an isolation efficiency
of $90\%$~($99\%$) for muons from \Zmm decays with $\pt=\SI[parse-numbers=false]{25 (60)}{\GeV}$.
Similarly to electrons, correction
factors are applied to muons to account for the small differences between data and simulation~\cite{PERF-2015-10}.
 
Jets are reconstructed from topological clusters of energy deposited
in the calorimeter~\cite{PERF-2014-07} using the anti-$k_t$
algorithm~\cite{antikt1} with a radius parameter of 0.4 (1.0) for
small-$R$ (large-$R$) jets, as implemented in the FastJet package~\cite{Fastjet}.
 
Small-$R$ jets are calibrated using an energy- and $\eta$-dependent
simulation-based calibration scheme with corrections derived from
data~\cite{ATL-PHYS-PUB-2015-015}. Jets are accepted within the
fiducial region $|\eta|<2.5$ and $\pt>\SI{30}{\GeV}$
($\pt>\SI{25}{\GeV}$) for the leptonic (hadronic) analysis. In the hadronic channel this threshold has been relaxed to increase forward-jet acceptance. Forward
jets in the region $2.5 <|\eta|<4.5$ are also considered in the
vector-like $T$-quark search analysis.  Quality criteria are imposed
to reject events that contain any jets arising from non-collision
sources or detector noise~\cite{ATLAS-CONF-2015-029}.
To reduce the contribution from jets
associated with pile-up, jets with $\pt<\SI{60}{\GeV}$ and $|\eta| <
2.4$ must satisfy a criterion that matches them to the hard-scatter vertex
using information from tracks reconstructed in the inner
tracking detector~\cite{ATLAS-CONF-2014-018}.
 
To prevent double counting of electron energy deposits as small-$R$
jets, the closest jet with distance $\Delta R_{y,\phi}\equiv\sqrt{{(\Delta y)}^2 +
{(\Delta\phi)}^2} < 0.2$ from a reconstructed electron
is removed. If the nearest surviving jet is within $\DeltaR_{y,\phi} = 0.4$ of the
electron, the electron is discarded to ensure it is sufficiently
separated from nearby jet activity. Jets with fewer than three tracks
and distance $\Delta R_{y,\phi} < 0.2$ from a muon are removed to reduce the
number of jet fakes from muons depositing energy in the
calorimeters. Muons with a distance $\Delta R_{y,\phi} < 0.4$ from any of the
surviving jets are removed to avoid contamination due to non-prompt
muons from heavy-flavour hadron decays.

Large-$R$ jets are trimmed~\cite{Krohn:2009th} to mitigate the impact of
initial-state radiation, underlying-event activity and pile-up. The
jet energy and pseudorapidity are further calibrated to account for
residual detector effects using energy- and $\eta$-dependent
calibration factors derived from simulation, with uncertainties
derived from data~\cite{ATLAS-CONF-2017-063}. Trimmed large-$R$ jets
are considered if they fulfil $\pt>\SI{250}{\GeV}$ and
$|\eta|<2.0$. To identify large-$R$ jets that are more likely to have
originated from hadronically decaying top-quarks than
from the fragmentation of other quarks and gluons, jet substructure
information is exploited.
 
In the trimming procedure, sub-jets, with radius \Rsub$=0.2$, are clustered starting from the
large-$R$ jet constituents using a $k_t$ algorithm. A sub-jet is
retained only if it contains at least $5\%$ of the total large-$R$ jet
transverse momentum, thereby removing the soft constituents from the
large-$R$ jet.
A top-tagging algorithm~\cite{ATL-PHYS-PUB-2015-053} is applied, corresponding to a loose working point with an approximately constant
top-tagging efficiency of $80\%$ above \pt of \SI{400}{\GeV}.
The algorithm depends on the calibrated jet mass,
measured from clusters in the calorimeter, and the
$N$-subjetiness ratio \tauthrtwo~\cite{Thaler:2010tr}.
The $N$-subjetiness \tauN~\cite{Thaler:2010tr} expresses how well a jet can be described as containing $N$ or fewer sub-jets.
The ratio \tauthrtwo = \tauthr/\tautwo allows discrimination between jets containing a three-prong structure and jets containing a two-prong structure.
 
In addition to calorimeter-based jets, jets reconstructed from inner
detector tracks using the anti-$k_t$ algorithm with a radius parameter
of \SI{0.2} are also used in the hadronic channel, following a similar strategy as in~\cite{Aaboud:2018mjh}.
They are referred to as track-based jets and are required to satisfy $\pt>\SI{10}{\GeV}$ and $|\eta|<2.5$.
 
Small-$R$ calorimeter-based and track-based jets with $|\eta|<2.5$ are
$b$-tagged as likely to contain $b$-hadrons using multivariate
techniques which exploit the long lifetime of $b$-hadrons and large
invariant mass of their decay products relative to $c$- and light hadrons~\cite{PERF-2012-04,ATL-PHYS-PUB-2016-012}. The
working point used provides an average tagging efficiency of $70\%$
for $b$-jets and a rejection factor of 12.2 (7.1) against
calorimeter-based (track-based) jets initiated by $c$-quarks and
381 (120) against calorimeter-based (track-based) jets initiated by
light-flavour quarks, in simulated \ttbar events. Correction factors
are derived and applied to correct for the small differences in
$b$-quark selection efficiency between data and MC
simulation~\cite{PERF-2012-04,ATLAS-CONF-2018-006,ATLAS-CONF-2018-001}.
 
The missing transverse momentum is calculated as the negative vector sum of the transverse momenta of particles in the event, and its magnitude is denoted \met. In
addition to the identified jets, electrons, muons, hadronically
decaying $\tau$-leptons and photons, a track-based soft term is
included in the \met calculation by considering tracks associated with
the hard-scattering vertex in the event which are not also associated
with an identified jet, electron, muon, hadronically decaying
$\tau$-lepton, or
photon~\cite{ATL-PHYS-PUB-2015-023,ATL-PHYS-PUB-2015-027}.
 
 
\section{Event selection and background estimation}
\label{sec:selection}
 
The experimental signature of mono-top events expected in the DM
(resonant and non-resonant) and vector-like $T$-quark models
considered is the presence of a top-quark and significant missing
transverse momentum, as seen in Section~\ref{sec:signalPheno}.
For the case of single VLT production, at least one additional forward jet is also expected.

The leptonic channel is only considered in order to target the non-resonant DM
model. In this model, the $u$-quark-initiated production of top-quarks
is favoured over anti-top-quark production, due to the PDF structure of the proton.
Therefore, positively charged leptons are favoured in the final state. Events that pass preselection are required to contain exactly one positively charged
lepton and one \btagged jet with $\pt>\SI{30}{\GeV}$. In order to
reduce the number of multijet background events, which are
characterised by low \met and low \Wboson~boson transverse
mass\footnote{The transverse mass of the lepton and \met system is
defined as $\mtW = \sqrt{2\pt(\ell)\met (1 - \cos \Delta \phi
(\pt(\ell), \met))}$, where $\pt(\ell)$ denotes the modulus of the
lepton transverse momentum, and $\Delta \phi (\pt(\ell), \met)$ the
azimuthal angle between the missing transverse momentum and the
lepton directions.} $\mtW$, it is also required that
$\met>\SI{50}{\GeV}$ and $\mtW+\met>\SI{60}{\GeV}$.
 
In the hadronic channel, because of the large expected Lorentz boost of the
top-quarks produced in the signal events, the top-quark decay products
can be collimated into a large-$R$ jet. This signature is used in both the non-resonant and
resonant DM models and the VLT models. Preselected events are then
required to contain zero leptons, one large-$R$ jet with
$\pt>\SI{250}{\GeV}$ and $|\eta|<2.0$. In order to suppress the multijet
background contribution, $\met>\SI{200}{\GeV}$ is also required.
 
\subsection{Signal region definition}

The signal region selection is optimised for the different considered benchmarks with simulated data, using variables tested and found to be well-modelled. In the optimisation the sensitivity is
estimated by performing a fit to the shape of the most discriminating observable including systematic uncertainties (see \Sect{\ref{sec:results}} for details). These observables are \met in the leptonic
channel and the transverse mass of the top-tagged large-$R$ jet ($J$) and the \met system, $m_{\text{T}}(\met,J)$~\footnote{The transverse mass of the large-$R$ jet and \met{} is defined as
$m_{\mathrm{T}}(\met,\mathrm{J}) = \sqrt{m(\mathrm{J})^{2} + 2\met \cdot \left(E_{\mathrm{T}(\mathrm{J})}-\pT(\mathrm{J})\cos (\Phi(\mathrm{J})-\Phi(\met))\right)}$, where $m_{\mathrm{T}}(\mathrm{J})$ is
the reconstructed invariant mass of the calibrated calorimeter-cluster constituents of a large-$R$ jet and $E_{\mathrm{T}}(\mathrm{J})$ is the projection of its energy in the transverse plane.}, in the
hadronic channel. For the tested mass hypothesis, the resulting best-performing selections lead to three signal regions: 1L-DM-SR for the non-resonant DM search in the leptonic channel and 0L-DM-SR and
0L-VLT-SR targeting the search in the hadronic channel for DM and VLT quarks, respectively.
 
In the leptonic channel, the mono-top signal is enhanced in regions
of phase space characterised by high $\mtW$ values. In addition, the
lepton and \btagged jet are closer to each other when originating from
the decay of a top-quark than in the case of  \Wjets and multijet
background events. Hence, in addition to the preselection described
previously, the region 1L-DM-SR is defined by requiring
$\mtW>\SI{260}{\GeV}$ and $\DeltaPhiLB < 1.2$.
 
In the hadronic channel, events in 0L-DM-SR and 0L-VLT-SR
are required to contain exactly one top-tagged large-$R$ jet with
$\pt>\SI{250}{\GeV}$ and one \btagged track-based jet, in addition to
the preselection criteria. The distance
between the top-tagged large-$R$ jet and the \met in the
transverse plane, $\Delta \Phi (\met,J)$, is required to fulfil
$\Delta \Phi (\met,J)\ge \pi /2$ since for signal events they are more
likely to be produced back-to-back. In order to suppress background
events due to fake \met mostly coming from jet mis-reconstruction in multijet
production, the asymmetry between \met and the \pt of the top-tagged
large-$R$ jet defined as $\Omega = (\MET - \pt(J))/(\MET + \pt(J))$
is required to be $\Omega > -0.3$. The multijet background is
additionally suppressed by requiring the minimum distance between
the \met and any small-$R$ jet in the transverse plane to be
$\Delta \Phi_\mathrm{min} > 1.0$. The signal region 0L-VLT-SR is
defined by requiring in addition at least one forward jet with
$\pt>\SI{25}{\GeV}$. The signal region requirements are summarised in \Tab{\ref{tab:RegionDef}}.
 
\subsection{Background estimation}
 
Dedicated control regions enriched in the dominant backgrounds are included in
the fit to constrain these backgrounds with data. Multijet
production background is estimated from data, while the rest of
background processes are taken from simulation.
 
The dominant background in the signal regions is due to \ttbar
production in both channels, representing 78\% of the total background
in the leptonic and 55\% (64\%) in the DM (VLT) hadronic channels.
This is followed by contributions from \Wjets (13\%) and single top
production (6.8\%) in the leptonic channel and from \Wjets and \Zjets
production, at the level of 12\% (13\%) for \Wjets and 14\% (15\%) for
\Zjets in the DM (VLT) signal regions, in the hadronic channel. A
minor background in the signal region with a non-negligible
contribution in the control regions is multijet production. The rest
of the backgrounds considered in the analysis are diboson production
as well as \ttbar production in association with a \Zboson, \Wboson or
Higgs boson.
 
The estimation of the multijet background is in particularly important
in the control regions used to estimate the main backgrounds. In the leptonic
channel the multijet background originates from either misidentification of a jet as a lepton candidate (fake lepton) or from
the presence of a non-prompt lepton (e.g., from a semileptonic $b$- or
$c$-hadron decay) that passes the isolation requirement. The shape and the normalisation
of the relevant distributions in multijet events and related systematic uncertainties are
estimated using a matrix method in the electron channel and the
anti-muon method in the muon channel~\cite{ATLAS-CONF-2014-058}.
The matrix method exploits differences in efficiencies to pass loose
or tight quality requirements~\cite{ATLAS-CONF-2016-024} between
prompt leptons, obtained from \Wboson and \Zboson decays, and non-prompt
or fake lepton candidates, from the misidentification of photons or jets.
These efficiencies are measured in dedicated control regions.
The prompt lepton efficiencies are
measured as a function of the \pt of the leading jet and the angular
distance between the lepton and its nearest jet, while the non-prompt
or fake efficiencies are parameterised in terms of the \pt of the
leading jet, the angle in the transverse plane between the lepton and
the \met and the \btagged jet multiplicity.  Multijet background
events containing non-prompt muons are modelled with the anti-muon method using a sample of
events enriched in non-isolated muons~\cite{ATLAS-CONF-2014-058}. Most
of these events originate from $b$- or $c$-hadron decays in
jets. These events pass the kinematic requirements of the selections
described in Section~\ref{sec:objects}. Only some of the muon identification criteria are
modified, ensuring there is no overlap with the signal selection. The
normalisation is determined using a binned maximum-likelihood fit to
the number of events observed in data in a control region dominated by
multijet events. This region is defined with the preselection criteria, but
removing the requirement on \met and requiring $\mtW <\SI{60}{\GeV}$.
 
In the hadronic channel the estimation of the multijet background is
performed using a set of control regions (B,C and D) dominated by
multijet background and defined to be orthogonal to the considered
signal region (0L-DM-SR or 0L-VLT-SR). The shape of the
multijet background is estimated from the control region B, which
differs from the signal region by requiring zero top-tagged large-$R$
jets. This shape is normalised by a factor which is calculated as the ratio of the
numbers of multijet events in regions C and D. Region C (region D) differs
from the signal region (B region) by requiring $\Delta \Phi (\met,J)
< \pi /2$.

In regions B and D, with zero top-tagged large-$R$
jets, $J$ is a large-$R$ jet chosen randomly from the selected large-$R$ jets. The multijet
contribution in these control regions is determined from the
difference between data and the residual contribution of other background
processes evaluated from simulation assuming the theoretical predictions for the corresponding
cross-sections.
 
\begin{table}[!ht]
\begin{center}
\begin{tabular}{l|cccc} \hline\hline
Selections (leptonic channel) &\multicolumn{2}{c}{1L-DM-SR} & 1L-TCR & 1L-WCR \\ \hline
Number of leptons & \multicolumn{2}{c}{= 1}   & = 1 & = 1 \\
$\pt(\ell)$ [\GeV] & \multicolumn{2}{c}{> 30} & > 30 & > 30 \\
Lepton charge & \multicolumn{2}{c}{> 0}  & > 0  & > 0   \\
Number of jets & \multicolumn{2}{c}{= 1} & = 2 & = 1  \\
Number of \btagged jets & \multicolumn{2}{c}{= 1} & = 2 & = 1  \\
$\pt$(\btagged jet) [\GeV] &  \multicolumn{2}{c}{> 30} & > 30 & > 30  \\
\met [\GeV] &\multicolumn{2}{c}{> 50} & > 50 & > 50 \\
$\mtW + \met$ [\GeV] & \multicolumn{2}{c}{> 60} & > 60 & > 60 \\
$\mtW$ [\GeV] &\multicolumn{2}{c}{> 260} &  $60 < \mtW < 100$ & $60 < \mtW < 100$ \\
$\DeltaPhiLB$ & \multicolumn{2}{c}{< 1.2} & -  & -  \\ \hline \hline
Selections (hadronic channel) & 0L-DM-SR & 0L-VLT-SR &
0L-TCR & 0L-VCR  \\ \hline
Number of forward jets & = 0 &  $\geq$ 1  & - & -  \\
Number of leptons  & \multicolumn{2}{c}{= 0} & = 0 & = 0 \\
\met [\GeV] & \multicolumn{2}{c}{$> \,$ 200} & $> \,$ 200 & $> \,$ 200 \\
Number of large-$R$ jets & \multicolumn{2}{c}{$\geq$ 1} & $\geq$ 1 & $\geq$ 1 \\
Number of top-tagged jets & \multicolumn{2}{c}{$\geq$ 1} & $\geq$ 1 & $\geq$ 1 \\
$\Delta \Phi(\met,\, J)$ & \multicolumn{2}{c}{$> \, \frac{\pi}{2}$} & $> \, \frac{\pi}{2}$ & $> \, \frac{\pi}{2}$ \\
Number of track-jets & \multicolumn{2}{c}{$\geq$ 1} & $\geq$ 1 & $\geq$ 1 \\
Number of \btagged track-jets & \multicolumn{2}{c}{= 1} & $\geq$ 2 & = 0 \\
Veto jet (masked tile-calo) & \multicolumn{2}{c}{-} & applied & - \\
$\Omega = \frac{\met - \pt(J)}{\met + \pt(J)}$ & \multicolumn{2}{c}{$> \,$ --0.3} & $> \,$ --0.3 & $> \,$ --0.3 \\
$\Delta \Phi_{\mathrm{min}}(\met,\, \mathrm{calo\, jets})$ & \multicolumn{2}{c}{$> \,$
1.0} & 0.2 $< \,$ $\Delta \Phi_{\mathrm{min}}$ $< \,$ 1.0 & $> \,$
1.0 \\
\hline\hline
\end{tabular}
\caption{Overview of the event selections used to define the signal
and control regions.}
\label{tab:RegionDef}
\end{center}
\end{table}
 
The control regions are defined to be orthogonal to each other and to the signal region.
They are required to fulfil the preselection criteria.  In the leptonic
channel, control regions enriched in \ttbar and \Wjets processes are
used (referred to as 1L-TCR and 1L-WCR, respectively). In
the hadronic channel, a control region enriched in \ttbar production
(referred to as 0L-TCR) and a region enriched in both the \Wjets
and \Zjets processes (referred to as 0L-VCR) are defined.
 
The control regions in the leptonic channel, 1L-TCR and
1L-WCR, are defined by modifying the requirement on $\mtW$ to a
window around the \Wboson mass, $60$~\GeV~$ < \mtW < 100 $~\GeV, and
removing the requirement on $\DeltaPhiLB$. For the 1L-TCR, events
are also required to contain a second \btagged jet.  The \ttbar
control region in the hadronic channel, 0L-TCR, is defined by
requiring two \btagged track-based jets and the minimum distance
between the \met and any small-$R$ jet in the transverse plane to
satisfy $\Delta \Phi_\mathrm{min} < 1$ (in order to reduce the signal
contribution) and $\Delta \Phi_\mathrm{min} > 0.2$ in order to suppress the
multijet background). Events with calorimeter-based jets located close to
disabled modules of the hadronic calorimeter are vetoed.
In the 0L-VCR control region a veto on \btagged track-based jets is
applied.
 
\begin{figure}[htbp]
\begin{center}
\begin{tabular}{cc}
\includegraphics[width=.49\textwidth]{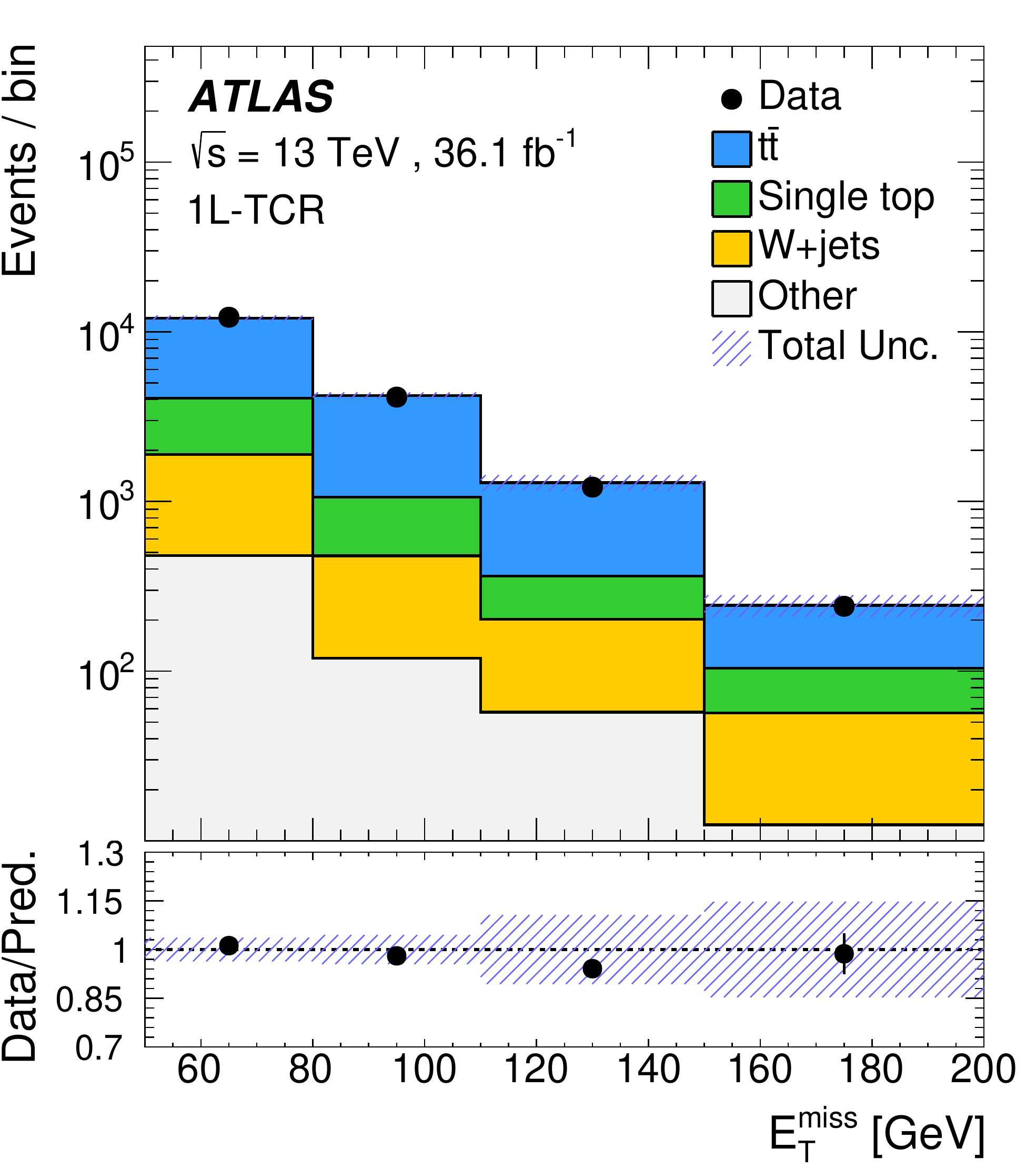} &
\includegraphics[width=.49\textwidth]{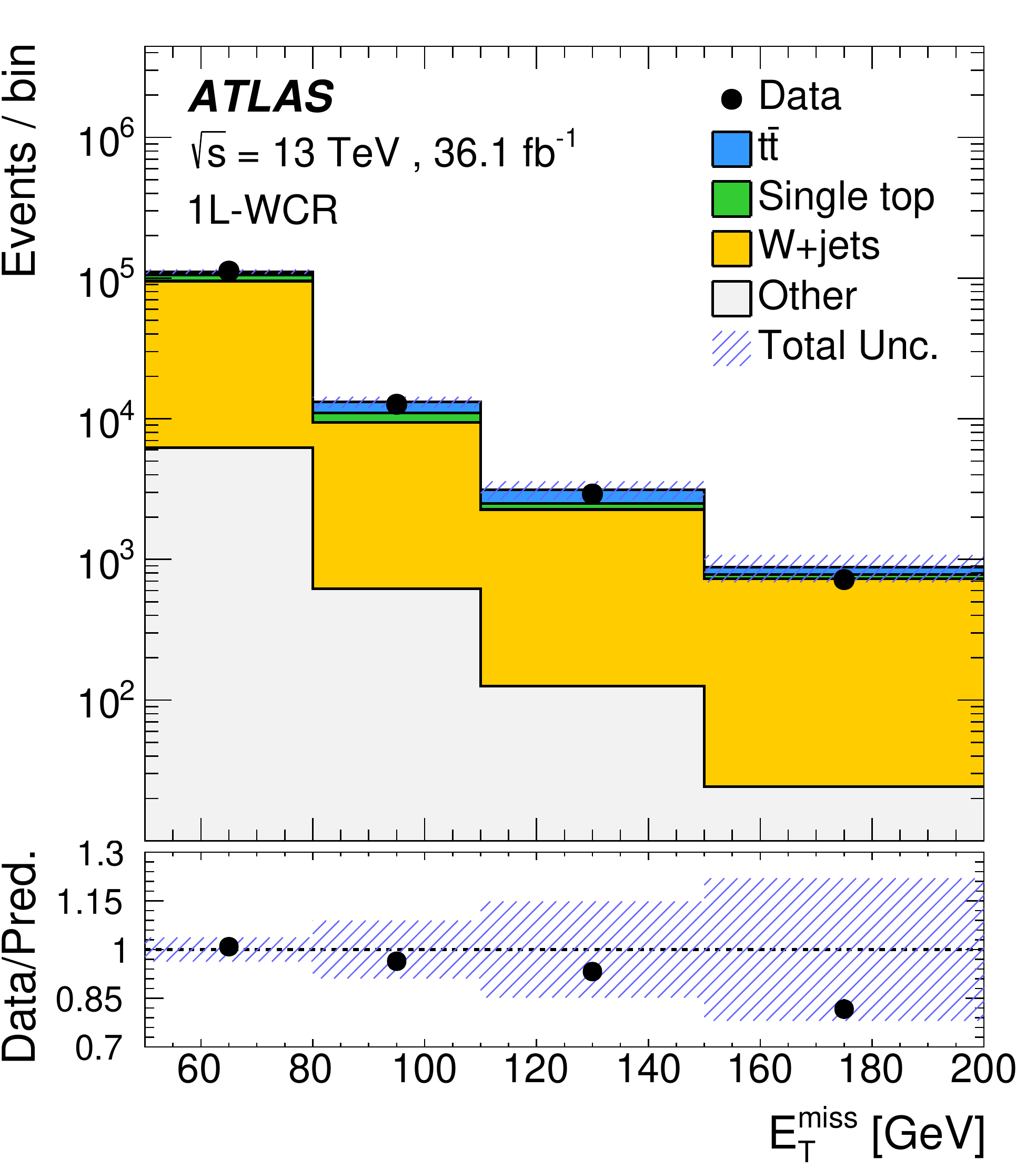} \\[-1mm]
(a) & (b) \\[2mm]
\includegraphics[width=.49\textwidth]{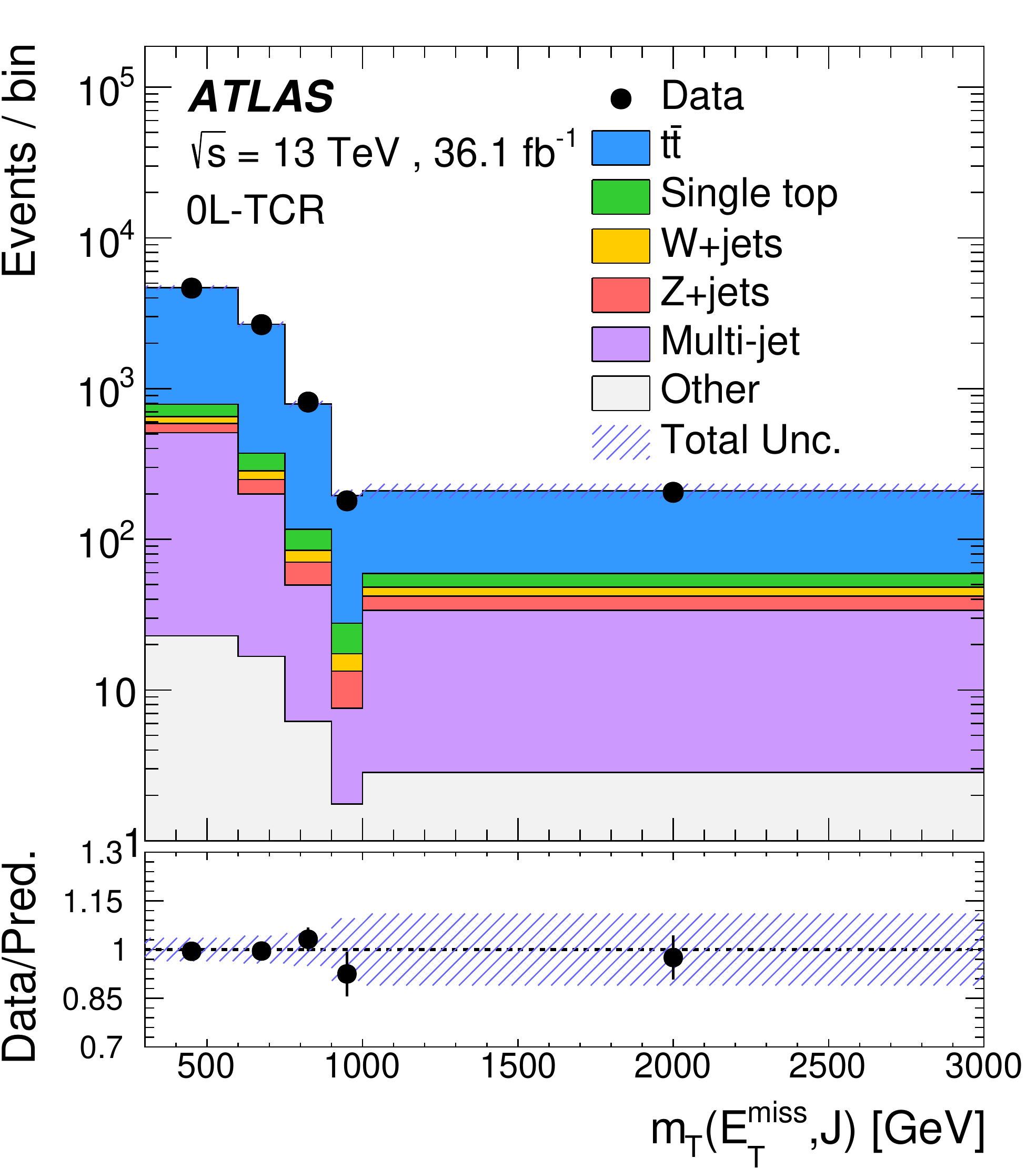} &
\includegraphics[width=.49\textwidth]{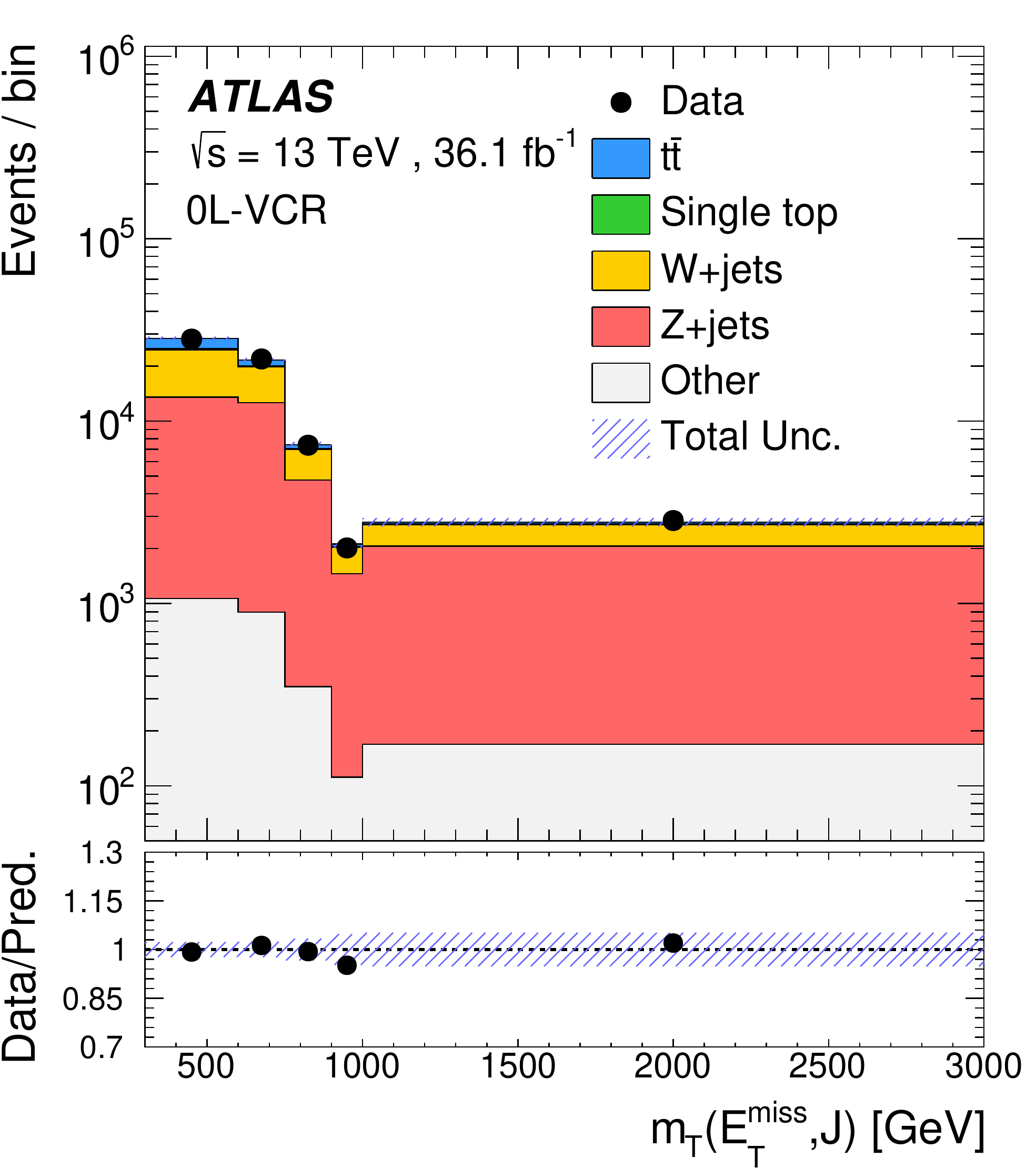} \\[-1mm]
(c) & (d) \\[2mm]
\end{tabular}
\end{center}
\caption{\label{fig:CRDataMC}
Comparison of data and SM prediction for the \met distribution in
(a) the \ttbar and (b) \Wjets control regions; and for the transverse mass
of the top-tagged large-$R$ jet and \met system,
$m_{\text{T}}(\met,J)$, distribution in the (c) \ttbar and (d)
\Wboson/\Zboson+jet control regions used for the dark-matter search ((a) and (b)) and
vector-like $T$-quark search ((c) and (d)). Other backgrounds in the 1L regions include multi-jet, $Z$+jets and diboson contributions, while in the 0L regions it is composed of diboson, \ttbar~+$X$ and multi-jet contributions.
The expectations in the leptonic (hadronic) channel are obtained
from a fit of the background-only hypothesis
to data in the 1L (0L) control regions, where the normalisations of the \ttbar and \Wjets (\ttbar
and \Wboson/\Zboson+\,jets) processes are treated as nuisance parameters in the fit.
The error bands include
statistical and systematic uncertainties. The last bin contains the
overflow events.
}
\end{figure}

\Tab{\ref{tab:RegionDef}} details the control region
selection in comparison with the signal region requirements.
A comparison of the observed and expected distributions for \met and
$m_{\text{T}}(\met,J)$ in the control regions is shown
in \Fig{\ref{fig:CRDataMC}} for the leptonic and hadronic channel,
respectively. The expectations in the leptonic (hadronic) channel are obtained
from a fit of the background-only hypothesis
to data in the 1L (0L) control regions, where the normalisations of the \ttbar and \Wjets (\ttbar
and \Wboson/\Zboson+\,jets) processes are treated as nuisance parameters in the fit (see \Sect{\ref{sec:results}}
for details of the fit).
 
\section{Systematic uncertainties}
\label{sec:systematic}

The normalisation and shapes of the signal and background estimates
are affected by systematic uncertainties from experimental sources
and theoretical predictions. Each source of uncertainty is
included as a nuisance parameter in the likelihood fit that determines
the possible signal contribution. The analysis is limited by
statistical uncertainties, thus the inclusion of systematic uncertainties leads
to only a small degradation of the expected sensitivity.
 
The sources of experimental uncertainty include the uncertainty in the
lepton trigger, identification and isolation efficiencies, the lepton energy and momentum scale and resolution~\cite{ATLAS-CONF-2016-024,PERF-2015-10,PERF-2014-07},
the \met{} trigger and
track-based soft-term scale and resolution~\cite{ATL-PHYS-PUB-2015-023,ATL-PHYS-PUB-2015-027}, the
jet pile-up rejection requirement, energy scale and resolution~\cite{PERF-2016-04},
resolutions for relevant large-$R$ jet properties (mass, transverse momentum and
the $N$-subjetiness ratio
\tauthrtwo), the $b$-tagging
efficiency~\cite{PERF-2012-04,ATL-PHYS-PUB-2016-012}, the pile-up
reweighting, and the luminosity.
 
The uncertainty in the combined 2015+2016 integrated luminosity is \lumiUncPct{\%},
derived following a methodology similar to that detailed in Ref.~\cite{DAPR-2013-01}, using a calibration of the
luminosity scale through $x$--$y$ beam-separation scans and using the LUCID-2 detector for
the baseline luminosity measurements \cite{LUCID2}. This systematic
uncertainty is applied to all backgrounds and signals that are estimated
using MC events, which are normalised to the measured
integrated luminosity.

Theoretical cross-section uncertainties are applied to the
normalisation of the simulated processes.
Additional shape uncertainties stemming from theoretical estimations are calculated by
comparing samples simulated with different assumptions and are
estimated for the dominant backgrounds.
 
Uncertainties in the modelling of the \ttbar and $t$-channel
single top background come from the choice of NLO-matching
method, the parton shower and hadronisation modelling, and the amount of
additional gluon radiation. The NLO-matching
uncertainty is estimated by comparing events produced with
\POWHEGBOX{} and \MGMCatNLO~\cite{Alwall:2014hca}, both interfaced
with \HERWIGpp~\cite{Bahr:2008pv}. The parton shower, hadronisation,
and underlying-event model uncertainty is estimated by comparing two
parton shower models, \PYTHIA{} and \HERWIGpp, while keeping the same
hard-scatter matrix element generator. Variations of the amount of
additional gluon radiation are estimated by comparing simulated
samples with enhanced or reduced radiation and different values of
tunable parameters related to additional
radiation~\cite{ATL-PHYS-PUB-2015-002}. The choice of scheme to
account for the interference between the $Wt$ and \ttbar processes
constitutes another source of systematic uncertainty that is
estimated by comparing samples using either the diagram removal scheme
or the diagram subtraction scheme~\cite{Frixione:2008yi}.
 
Modelling uncertainties affecting the shape of the
\Wboson/\Zboson+\,jets background are estimated in the hadronic
channel, where these processes constitute an important background. An
uncertainty in the modelling of \Wboson/\Zboson+\,jets is estimated by
comparing the nominal simulation with a \MGMCatNLO simulation in which
matrix elements were calculated at LO for up to four partons. In
addition, the effects of independently varying the scales for the
renormalisation, factorisation, and resummation by factors of 0.5 and
2 are used. Since the \Wboson/\Zboson+\,jets background is constrained
by a control region with a veto on \btagged jets, an additional
uncertainty related to the $b$-flavour content in the
\Wboson/\Zboson+\,jets background is taken into account by varying the
number of events containing $b$-hadrons by
50\%~\cite{Aaboud:2017yqf,Aad:2013vka}.  Uncertainties in the
modelling of the signal samples have been evaluated for signal points
close to the expected exclusion mass limits and found to be
negligible.
 
The effects of parton distribution function (PDF) uncertainties on the acceptance of the \ttbar
and \Wboson/\Zboson+\,jets backgrounds are estimated following the PDF4LHC
prescription~\cite{Butterworth:2015oua}.
 
The systematic uncertainty of 50\%{} associated with the data-driven
modelling of the multijet events is estimated in the leptonic channel,
based on comparisons of the rates obtained using alternative methods,
as described in previous analyses~\cite{ATLAS-CONF-2014-058}. In the
case of the hadronic channel, this systematic is derived from a
closure test of the data-driven method in a multijet-dominated
validation region using simulated dijet samples.
 
A breakdown of the effects of the various sources of systematic
uncertainty on the background prediction is presented
in \Tab{\ref{tab:SystematicsDM}} and \Tab{\ref{tab:SystematicsVLT}} for the two searches. The relative effects on the background yields in the signal region after the simultaneous fit to data  in the signal and control regions are shown. The dominant background modelling uncertainties are due to the modelling of single top $Wt$ production for the leptonic channel and the modelling of the $b$-flavour content in the \Wboson/\Zboson+\,jets backgrounds.
 
\begin{table}[htbp]
\begin{center}
\begin{tabular}{l|cc|cccccc}
 
\hline \hline
\multirow{2}{*}{}&\multicolumn{2}{c}{1L-DM-SR}\vline&\multicolumn{6}{c}{0L-DM-SR}\\
 
\cline{2-9}
& non-$t\bar{t}$  & $t\bar{t}$ & $t\bar{t}$      & Single top      & $W$+jets      & $Z$+jets      & Multijet      & Other \\
\hline
$b$-tagging &  \num[round-mode=places,round-precision=1]{4.77}  &  \num[round-mode=places,round-precision=1]{4.61}  &  \num[round-mode=places,round-precision=1]{4.11} &  \num[round-mode=places,round-precision=1]{2.94} &  \num[round-mode=places,round-precision=1]{9.24} &  \num[round-mode=places,round-precision=1]{7.73} & - &  \num[round-mode=places,round-precision=1]{7.99}  \\
\met &  \num[round-mode=places,round-precision=0]{11.7}\hphantom{.00}  &  \num[round-mode=places,round-precision=1]{1.14}  & \num[round-mode=places,round-precision=1]{2.2} &  \num[round-mode=places,round-precision=1]{2.12} &  \num[round-mode=places,round-precision=1]{2.2} &  \num[round-mode=places,round-precision=1]{2.16} &  \num[round-mode=places,round-precision=1]{2.99} &  \num[round-mode=places,round-precision=1]{1.99}\\
Large-$R$ jets &  - & - & \num[round-mode=places,round-precision=1]{9.04} &  \num[round-mode=places,round-precision=1]{9.5} &  \num[round-mode=places,round-precision=0]{13.2}\hphantom{.00} &  \num[round-mode=places,round-precision=0]{12.5}\hphantom{.00} & - &  \num[round-mode=places,round-precision=0]{11.8}\hphantom{.00} \\
Small-$R$ jets &  \num[round-mode=places,round-precision=1]{9.9}  &  \num[round-mode=places,round-precision=1]{6.96} & \num[round-mode=places,round-precision=1]{1.27} &  \num[round-mode=places,round-precision=1]{2.93} &  \num[round-mode=places,round-precision=1]{1.01} &  \num[round-mode=places,round-precision=1]{0.511} & - &  \num[round-mode=places,round-precision=1]{0.968} \\
Lepton &  \num[round-mode=places,round-precision=1]{1.19}  &  \num[round-mode=places,round-precision=1]{0.752} & < 0.1\hphantom{ <} & < 0.1\hphantom{ <} & < 0.1\hphantom{ <} & < 0.1\hphantom{ <} & - & < 0.1\hphantom{ <} \\
Luminosity &  \num[round-mode=places,round-precision=1]{2.03} &  \num[round-mode=places,round-precision=1]{2.09}  & \num[round-mode=places,round-precision=1]{2.122} &  \num[round-mode=places,round-precision=1]{2.11} &  \num[round-mode=places,round-precision=1]{2.25} &  \num[round-mode=places,round-precision=1]{2.22} & - &  \num[round-mode=places,round-precision=1]{2.07}  \\
Pile-up  &  \num[round-mode=places,round-precision=1]{5.29}  &  \num[round-mode=places,round-precision=1]{1.43} & \num[round-mode=places,round-precision=1]{0.268} &  \num[round-mode=places,round-precision=1]{0.346} &  \num[round-mode=places,round-precision=1]{0.818} &  \num[round-mode=places,round-precision=1]{1.23} & - &  \num[round-mode=places,round-precision=1]{1.37} \\
Background modelling &  \num[round-mode=places,round-precision=0]{14.8}\hphantom{.00}  &  \num[round-mode=places,round-precision=0]{13.5}\hphantom{.00} & \num[round-mode=places,round-precision=1]{8.85} &  \num[round-mode=places,round-precision=1]{5.28} &  \num[round-mode=places,round-precision=0]{26.9}\hphantom{.00} &  \num[round-mode=places,round-precision=0]{26.9}\hphantom{.00} &  \num[round-mode=figures,round-precision=2]{111}\hphantom{.000} &  \num[round-mode=places,round-precision=1]{1.04}\\
\hline
Total systematic &  \num[round-mode=places,round-precision=0]{18.2}\hphantom{.00}  &  \num[round-mode=places,round-precision=0]{11.8}\hphantom{.00} & \num[round-mode=places,round-precision=1]{6.76} &  \num[round-mode=places,round-precision=0]{13.2}\hphantom{.00} &  \num[round-mode=places,round-precision=0]{31.5}\hphantom{.00} &  \num[round-mode=places,round-precision=0]{30.5}\hphantom{.00} &  \num[round-mode=places,round-precision=0]{88.5}\hphantom{.00} &  \num[round-mode=places,round-precision=0]{16.2}\hphantom{.00}\\
\hline \hline
\end{tabular}
\caption{Relative effect (in \%) of various sources of systematic uncertainty on the predicted background yields in the signal regions used for the dark-matter search, obtained after the fit to data.
Individual sources of uncertainties are correlated, and their sum in quadrature is not necessarily equal to the total background uncertainty.}
\label{tab:SystematicsDM}
\end{center}
\end{table}

\begin{table}[htbp]
\begin{center}
\begin{tabular}{l|ccccc}
\hline \hline
\multirow{2}{*}{}&\multicolumn{5}{c}{0L-VLT-SR}\\
 
\cline{2-6}
& $t\bar{t}$      & Single top      & $W$+jets      & $Z$+jets      & Other \\
\hline
$b$-tagging & \num[round-mode=places,round-precision=1]{3.93} &  \num[round-mode=places,round-precision=1]{3.16} &  \num[round-mode=places,round-precision=0]{11.0}\hphantom{.00} &  \num[round-mode=places,round-precision=1]{7.69} &  \num[round-mode=places,round-precision=1]{8.4}  \\
\met & \num[round-mode=places,round-precision=1]{1.41} &  \num[round-mode=places,round-precision=1]{1.86} &  \num[round-mode=places,
round-precision=1]{0.776} &  \num[round-mode=places,round-precision=1]{0.525} &  \num[round-mode=places,round-precision=1]{0.25}\\
Large-$R$ jets & \num[round-mode=places,round-precision=0]{10.6}\hphantom{.00} &  \num[round-mode=places,round-precision=0]{12.4}\hphantom{.00} &  \num[round-mode=places,round-precision=0]{14.7}\hphantom{.00} &  \num[round-mode=places,round-precision=0]{12.9}\hphantom{.00} &  \num[round-mode=places,round-precision=0]{13.3}\hphantom{.00} \\
Small-$R$ jets & \num[round-mode=places,round-precision=1]{7.3} &  \num[round-mode=places,round-precision=1]{7.79} &  \num[round-mode=places,round-precision=1]{7.61} &  \num[round-mode=places,round-precision=1]{8.29} &  \num[round-mode=places,round-precision=0]{10.9}\hphantom{.00} \\
Lepton & < 0.1\hphantom{ <} & < 0.1\hphantom{ <} & < 0.1\hphantom{ <} & < 0.1\hphantom{ <} & < 0.1\hphantom{ <} \\
Luminosity & \num[round-mode=places,round-precision=1]{2.18} &  \num[round-mode=places,round-precision=1]{2.29} &  \num[round-mode=places,round-precision=1]{2.40} &  \num[round-mode=places,round-precision=1]{2.12} &  \num[round-mode=places,round-precision=1]{2.24} \\
Pile-up  & \num[round-mode=places,round-precision=1]{1.27} &  \num[round-mode=places,round-precision=1]{2.91} &  \num[round-mode=places,round-precision=1]{4.65} &  \num[round-mode=places,round-precision=1]{4.09} &  \num[round-mode=places,round-precision=1]{5.12} \\
Background modelling & \num[round-mode=places,round-precision=0]{14.2}\hphantom{.00} &  \num[round-mode=places,round-precision=1]{6.06} &  \num[round-mode=places,round-precision=0]{28.4}\hphantom{.00} &  \num[round-mode=places,round-precision=0]{26.4}\hphantom{.00} &  \num[round-mode=places,round-precision=1]{2.68}\\\hline
Total systematic & \num[round-mode=places,round-precision=1]{7.89} & \num[round-mode=places,round-precision=1]{15.5}\hphantom{0} & \num[round-mode=places,round-precision=0]{21.3}\hphantom{.00} & \num[round-mode=places,round-precision=0]{19.4}\hphantom{.00} & \num[round-mode=places,round-precision=1]{19.5}\hphantom{0} \\
\hline \hline
\end{tabular}
\caption{Relative effect (in \%) of various sources of systematic uncertainty on the predicted background yields in the signal region used for the vector-like $T$-quark search, obtained after the fit to
data.
Individual sources of uncertainties are correlated, and their sum in quadrature is not necessarily equal to the total background uncertainty.}
\label{tab:SystematicsVLT}
\end{center}
\end{table}
 
\section{Results}
\label{sec:results}
 
In order to test for the presence of a signal, a simultaneous fit to data
in the signal and control regions is performed. The fit is based on a
profile-likelihood technique, where systematic uncertainties are
allowed to vary as Gaussian-distributed nuisance parameters (NP)
and subsequently acquire their best-fit values. Additionally, the dominant backgrounds are constrained
by treating their normalisation as NP in the fit.  The calculation of confidence
intervals and hypothesis testing is performed using a frequentist method as implemented in
RooStats~\cite{RooFit} using the asymptotic approximation~\cite{Cowan:2010js}.
 
The \met distribution is used in the 1L signal region and the number of
events is used instead in the control regions, while for the case of the 0L regions
the distribution of the transverse mass of the top-tagged large-$R$ jet
and \met system, $m_{\text{T}}(\met,J)$, is used in signal and control regions. For each of
the three fits the binning of the distributions is optimised
separately to obtain the highest expected sensitivity. For the testing of the
non-resonant DM signal, both the 1L and 0L regions
are used simultaneously in the fit (two signal regions and four
control regions).  For the resonant DM and VLT tests the fits
are performed in the corresponding 0L regions, one signal region
and two control regions for each fit. Uncertainties due to the limited
size of the simulated samples are taken into account in each bin of
the fitted distributions. Nuisance parameters accounting for
systematic uncertainties are not considered in the fit if they have an
impact on either normalisation or shape which is below 1\%. The systematic
uncertainties are symmetrised and also smoothed if the bin-to-bin
statistical variation is significant. Most uncertainties are found
to be neither significantly constrained nor pulled from their initial
values. Small variations are observed in the \ttbar modelling and
multijet background uncertainties due to the mis-modelling observed in
the shape of the transverse momentum distribution of
top-quarks~\cite{Aaboud:2017fha,Aaboud:2018eqg}.
Small variations are also observed in the large-$R$
jet and \Zboson/\Wboson+\,jets modelling uncertainties.
The results of the fit show that the data are
compatible with the background-only hypothesis.
 
\begin{figure}[htbp]
\begin{center}
\begin{tabular}{cc}
\includegraphics[width=.49\textwidth]{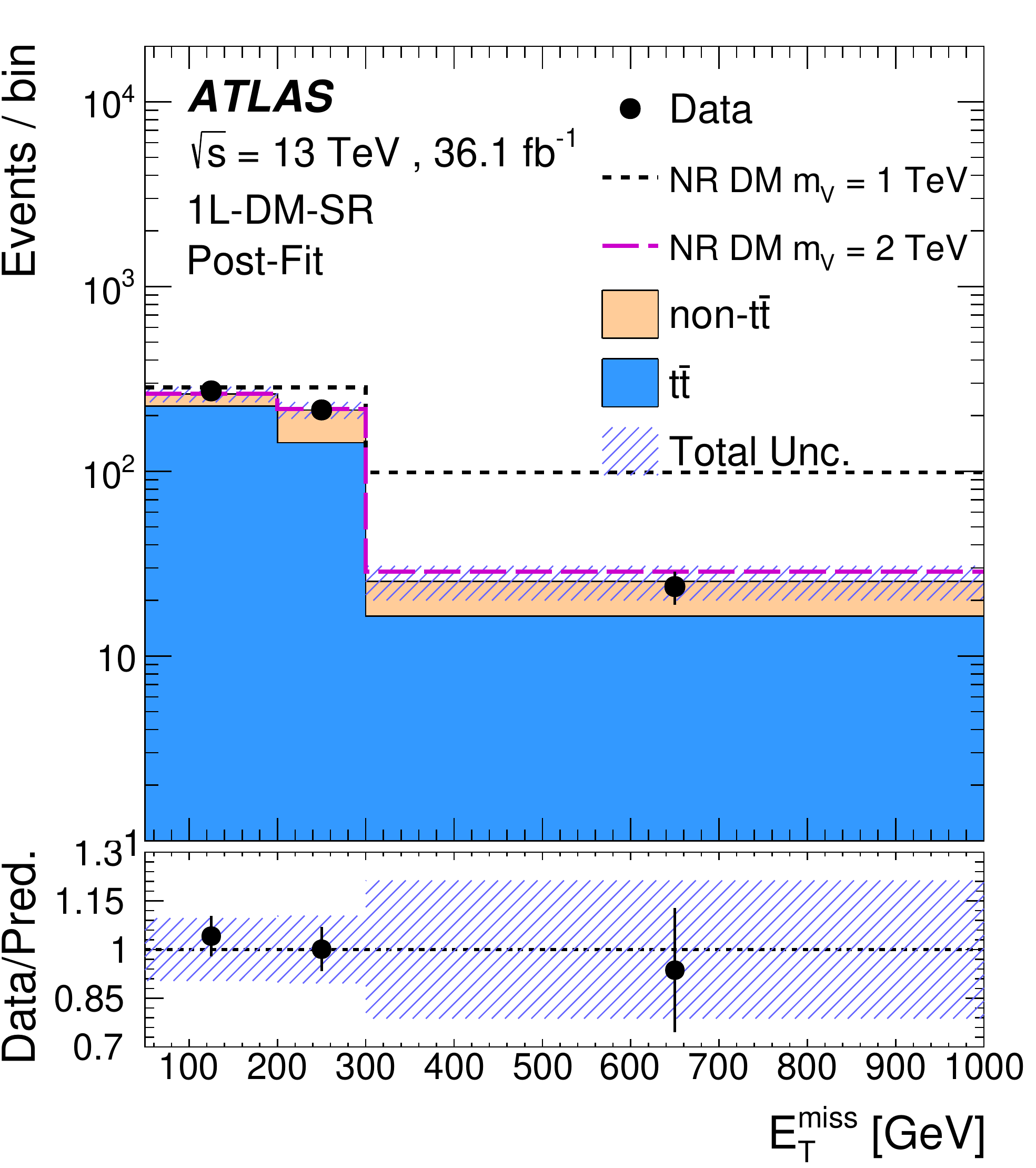} &
\includegraphics[width=.49\textwidth]{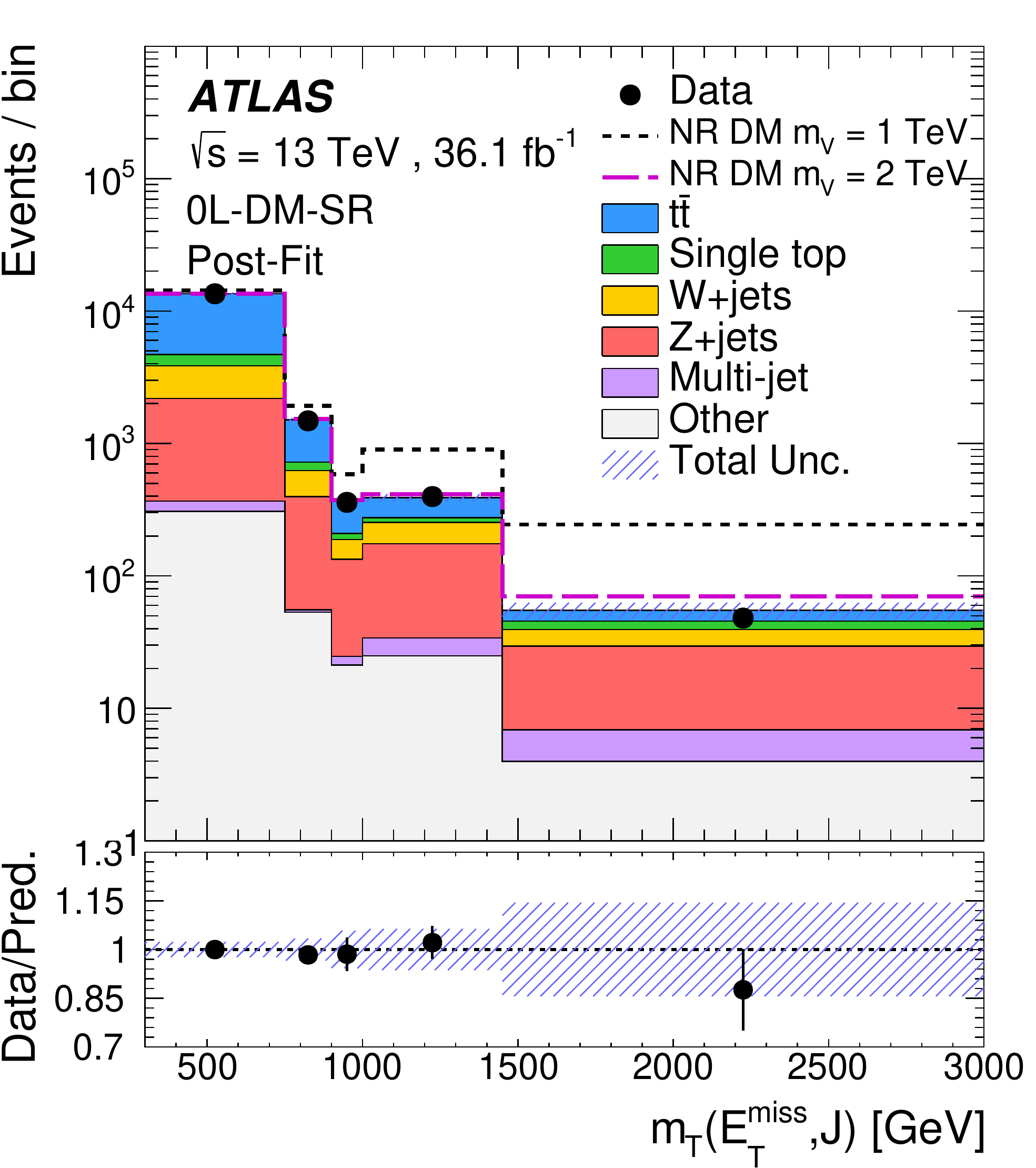} \\[-1mm]
(a) & (b) \\[2mm]
\includegraphics[width=.49\textwidth]{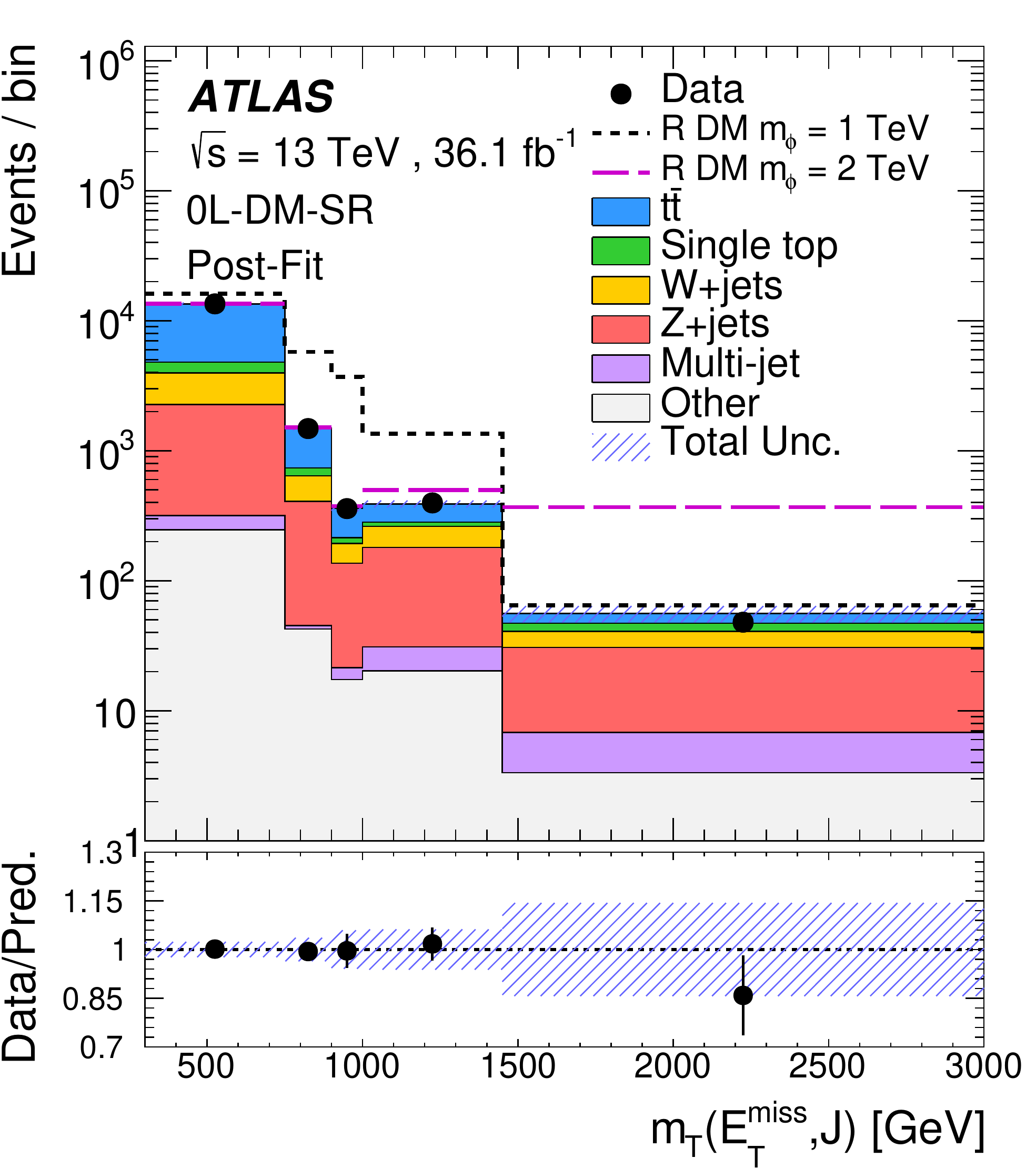} &
\includegraphics[width=.49\textwidth]{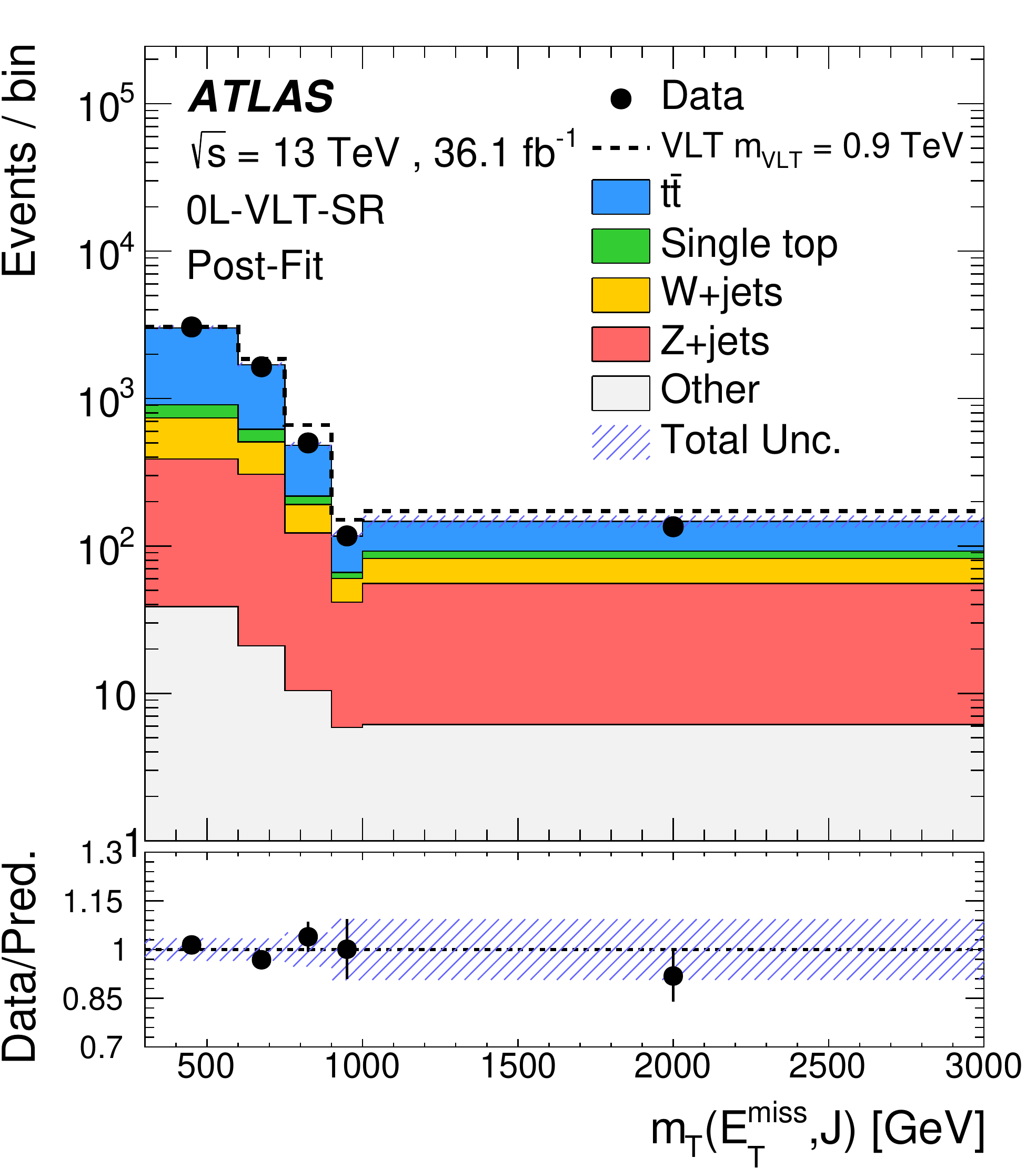} \\[-1mm]
(a) & (b) \\
\end{tabular}
\end{center}
\caption{\label{fig:DistributionFitted}
Comparison of data and fitted expectations for the \met and the transverse mass of
the top-tagged large-$R$ jet and \met system, $m_{\text{T}}(\met,J)$,
distributions in the signal regions.  Other backgrounds in the 1L regions include multi-jet, $Z$+jets and diboson contributions, while in the 0L regions it is composed of diboson and \ttbar~+$X$ contributions. The background-only hypothesis is used in the
fit: (a) and (b) including the 1L and 0L DM signal regions as well as the
1L and 0L control regions; (c) 0L DM signal and control
regions; (d) 0L VLT signal and control regions. The error bands
include statistical and systematic uncertainties. The expected shape of a
benchmark signal normalised to the theoretical prediction is added on top of the
SM prediction. The benchmark signals correspond to: the non-resonant (NR) DM model
with $m_{V}=1$~\TeV\ and 2~\TeV, $m_\chi=1$~\GeV, $a=0.5$ and $g_\chi=1$; the resonant
(R) DM model with $m_{\phi}=1$~\TeV\ and 2~\TeV, $m_\chi=10$~\GeV, $\lambda = 0.2$ and
$y=0.4$; and a VLT with a mass of 0.9~\TeV.}
 
\end{figure}

The numbers of events observed in the signal and control regions are
presented in \Tab{\ref{tab:YieldsFitted}}, together with the backgrounds estimated
prior to the simultaneous fit. The distribution of the observable
used in the fit (\met or $m_{\text{T}}(\met,J)$) in the signal regions
for data and the fitted SM expectation under the background-only
hypothesis are shown in \Fig{\ref{fig:DistributionFitted}}. In these plots,
the expected contribution from a benchmark signal is also shown for comparison.  No
significant excess above the SM expectation is found in any of the signal regions.
 
\begin{table}[htbp]
\begin{center}
\resizebox{\textwidth}{!}{
\begin{tabular}{l |r@{~$\pm$~}l  r@{~$\pm$~}l  r@{~$\pm$~}l r@{~$\pm$~}l r@{~$\pm$~}l r@{~$\pm$~}l r@{~$\pm$~}l}
\hline\hline
& \multicolumn{2}{c}{1L-DM-SR} &\multicolumn{2}{c}{1L-TCR} & \multicolumn{2}{c}{1L-WCR} & \multicolumn{2}{c}{0L-DM-SR} &
\multicolumn{2}{c}{0L-VLT-SR} & \multicolumn{2}{c}{0L-TCR} & \multicolumn{2}{c}{0L-VCR}\\
\hline
$t\bar{t}$ &\num[round-mode=figures,round-precision=2]{391.078}
& \num[round-mode=figures,round-precision=2]{138.203}
& \num[round-mode=figures,round-precision=3]{12308.2}
& \num[round-mode=figures,round-precision=2]{3097.94}
& \num[round-mode=figures,round-precision=2]{8432.74}
& \num[round-mode=figures,round-precision=2]{1660.64}
& \num[round-mode=figures,round-precision=3]{10187.5}
& \num[round-mode=figures,round-precision=2]{2851.74}
& \num[round-mode=figures,round-precision=2]{3746.14}
& \num[round-mode=figures,round-precision=2]{1231.66} & \num[round-mode=figures,round-precision=2]{6979.81} &  \num[round-mode=figures,round-precision=2]{1721.48} & \num[round-mode=figures,round-precision=2]{6067.23} & \num[round-mode=figures,round-precision=2]{1834.51} \\
Single top  &\num[round-mode=figures,round-precision=2]{65.6128}
& \num[round-mode=figures,round-precision=2]{20.7592}
& \num[round-mode=figures,round-precision=3]{2932.23}
& \num[round-mode=figures,round-precision=2]{760.84}
& \num[round-mode=figures,round-precision=3]{12174.8}
& \num[round-mode=figures,round-precision=2]{1725.53}
& \num[round-mode=figures,round-precision=3]{1015.82}
& \num[round-mode=figures,round-precision=2]{258.631}
& \num[round-mode=figures,round-precision=3]{355.75}
& \num[round-mode=figures,round-precision=2]{96.6554} & \num[round-mode=figures,round-precision=3]{273.875} & \num[round-mode=figures,round-precision=2]{70.6229} & \num[round-mode=figures,round-precision=2]{888.052} & \num[round-mode=figures,round-precision=2]{254.136} \\
$W$+jets  &\num[round-mode=figures,round-precision=3]{34.2304}
& \num[round-mode=figures,round-precision=2]{8.40418}
& \num[round-mode=figures,round-precision=3]{1885.59}
& \num[round-mode=figures,round-precision=2]{640.723}
& \num[round-mode=figures,round-precision=2]{92109.8}
& \num[round-mode=figures,round-precision=2]{24086.8}
& \num[round-mode=figures,round-precision=3]{2239.95}
& \num[round-mode=figures,round-precision=2]{897.389}
& \num[round-mode=figures,round-precision=2]{769.55}
& \num[round-mode=figures,round-precision=2]{306.128} & \num[round-mode=figures,round-precision=3]{147.147} & \num[round-mode=figures,round-precision=2]{87.4522} & \num[round-mode=figures,round-precision=2]{28470.4} & \num[round-mode=figures,round-precision=2]{12134.7} \\
$Z$+jets &\num[round-mode=figures,round-precision=2]{0.398342}
& \num[round-mode=figures,round-precision=2]{0.858914}
& \num[round-mode=figures,round-precision=3]{112.085}
& \num[round-mode=figures,round-precision=2]{48.9708}
& \num[round-mode=figures,round-precision=3]{3406.78}
& \num[round-mode=figures,round-precision=2]{985.647}
& \num[round-mode=figures,round-precision=2]{2655.56}
& \num[round-mode=figures,round-precision=2]{1125.67}
& \num[round-mode=figures,round-precision=2]{849.96}
& \num[round-mode=figures,round-precision=2]{361.143} & \num[round-mode=figures,round-precision=3]{138.732} & \num[round-mode=figures,round-precision=2]{82.8244} & \num[round-mode=figures,round-precision=2]{27438} & \num[round-mode=figures,round-precision=2]{11043.4} \\
Other  &\num[round-mode=figures,round-precision=2]{13.8398}
& \num[round-mode=figures,round-precision=2]{14.9265}
& \num[round-mode=figures,round-precision=2]{643.225}
& \num[round-mode=figures,round-precision=2]{883.177}
& \num[round-mode=figures,round-precision=1]{6672.99}
& \num[round-mode=figures,round-precision=2]{10145.2}
& \num[round-mode=figures,round-precision=2]{531.821}
& \num[round-mode=figures,round-precision=2]{194.494}
& \num[round-mode=figures,round-precision=2]{88.81}
& \num[round-mode=figures,round-precision=2]{28.03}
& \num[round-mode=figures,round-precision=3]{1055.42}
& \num[round-mode=figures,round-precision=2]{638.27}
& \num[round-mode=figures,round-precision=3]{2731.51}
& \num[round-mode=figures,round-precision=2]{763.502} \\
\hline
Total Background & \num[round-mode=figures,round-precision=2]{501.000}
& \num[round-mode=figures,round-precision=2]{140.616}
& \num[round-mode=figures,round-precision=3]{17882.4}
& \num[round-mode=figures,round-precision=2]{3371.7998}
& \num[round-mode=figures,round-precision=3]{122818}
& \num[round-mode=figures,round-precision=2]{26262.108}
& \num[round-mode=figures,round-precision=3]{16560.6}
& \num[round-mode=figures,round-precision=2]{4507.61}
& \num[round-mode=figures,round-precision=2]{5810.21}
& \num[round-mode=figures,round-precision=2]{1733.06} & \num[round-mode=figures,round-precision=2]{8565.79} & \num[round-mode=figures,round-precision=2]{1960.04} & \num[round-mode=figures,round-precision=2]{65554.8} & \num[round-mode=figures,round-precision=2]{22253} \\
\hline
Data   & \multicolumn{2}{c}{511}\hphantom{$\,\pm\,$000} & \multicolumn{2}{c}{17$\,$662}\hphantom{$\,\pm\,$00$\,$000} & \multicolumn{2}{c}{127$\,$286}\hphantom{$\,\pm\,$000$\,$000} & \multicolumn{2}{c}{15$\,$781}\hphantom{$\,\pm\,$00$\,$000} & \multicolumn{2}{c}{5454}\hphantom{$\,\pm\,$0000} & \multicolumn{2}{c}{8493}\hphantom{$\,\pm\,$0000} & \multicolumn{2}{c}{62$\,$304}\hphantom{$\,\pm\,$00$\,$000} \\
\hline
R DM $m_{\phi}$ = 1 \TeV & \multicolumn{2}{c}{-} & \multicolumn{2}{c}{-} & \multicolumn{2}{c}{-} & \num[round-mode=figures,round-precision=3]{11260.8} & \num[round-mode=figures,round-precision=2]{1254.55} & \multicolumn{2}{c}{-} & \num[round-mode=figures,round-precision=2]{55.803} & \num[round-mode=figures,round-precision=2]{12.6643} & \num[round-mode=figures,round-precision=2]{8084.18} & \num[round-mode=figures,round-precision=2]{1579.65} \\
R DM $m_{\phi}$ = 2 \TeV & \multicolumn{2}{c}{-} & \multicolumn{2}{c}{-} & \multicolumn{2}{c}{-} & \num[round-mode=figures,round-precision=3]{468.977} & \num[round-mode=figures,round-precision=2]{83.2669} & \multicolumn{2}{c}{-} & \num[round-mode=figures,round-precision=2]{4.253} & \num[round-mode=figures,round-precision=2]{1.052} & \num[round-mode=figures,round-precision=3]{348.52} & \num[round-mode=figures,round-precision=2]{86.1307} \\\hline
NR DM $m_{\phi}$ = 1 \TeV & \num[round-mode=figures,round-precision=3]{164.818} & \num[round-mode=figures,round-precision=2]{22.94} & \num[round-mode=figures,round-precision=3]{1.019} & \num[round-mode=figures,round-precision=2]{0.465} & \num[round-mode=figures,round-precision=3]{20.18} & \num[round-mode=figures,round-precision=2]{2.77}  & \num[round-mode=figures,round-precision=3]{2090.29} & \num[round-mode=figures,round-precision=2]{283.251} & \multicolumn{2}{c}{-} & \num[round-mode=figures,round-precision=3]{28.9993} & \num[round-mode=figures,round-precision=2]{5.86888} & \num[round-mode=figures,round-precision=3]{1600.2} & \num[round-mode=figures,round-precision=2]{316.645} \\
NR DM $m_{\phi}$ = 2 \TeV & \num[round-mode=figures,round-precision=2]{6.49} & \num[round-mode=figures,round-precision=2]{2.66} & \num[round-mode=figures,round-precision=2]{0.0265} & \num[round-mode=figures,round-precision=2]{0.013} & \num[round-mode=figures,round-precision=3]{0.496} & \num[round-mode=figures,round-precision=2]{0.0965} & \num[round-mode=figures,round-precision=2]{94.6664} & \num[round-mode=figures,round-precision=2]{13.2137} & \multicolumn{2}{c}{-} & \num[round-mode=figures,round-precision=3]{1.07578} & \num[round-mode=figures,round-precision=2]{0.211743} & \num[round-mode=figures,round-precision=2]{75.178} & \num[round-mode=figures,round-precision=2]{15.2028} \\\hline
VLT $m_{\mathrm{VLT}}$ = 0.9 \TeV & \multicolumn{2}{c}{-} & \multicolumn{2}{c}{-} & \multicolumn{2}{c}{-} & \multicolumn{2}{c}{-} & \num[round-mode=figures,round-precision=3]{111.592} & \num[round-mode=figures,round-precision=2]{20.0503} & \num[round-mode=figures,round-precision=3]{21.011} & \num[round-mode=figures,round-precision=2]{5.331} & \num[round-mode=figures,round-precision=2]{76.1046} & \num[round-mode=figures,round-precision=2]{16.7274} \\
\hline \hline
\end{tabular}}
\caption{Numbers of events observed in the signal and control regions,
together with the estimated SM backgrounds before
the fit to data. The uncertainties include statistical and
systematic uncertainties.
The expected numbers of events for benchmark signals normalised to the theoretical prediction are also shown. The benchmark signals correspond to: the non-resonant (NR) DM model with $m_{V}=1$~\TeV\ and 2~\TeV, $m_\chi=1$~\GeV, $a=0.5$ and $g_\chi=1$; the resonant (R) DM model with $m_{\phi}=1$~\TeV\ and 2~\TeV, $m_\chi=10$~\GeV, $\lambda = 0.2$ and $y=0.4$; and a VLT with a mass of 0.9~\TeV.}
\label{tab:YieldsFitted}
\end{center}
\end{table}

\clearpage
 
\begin{figure}[htbp]
\begin{center}
\begin{tabular}{c c}
\includegraphics[width=.49\textwidth]{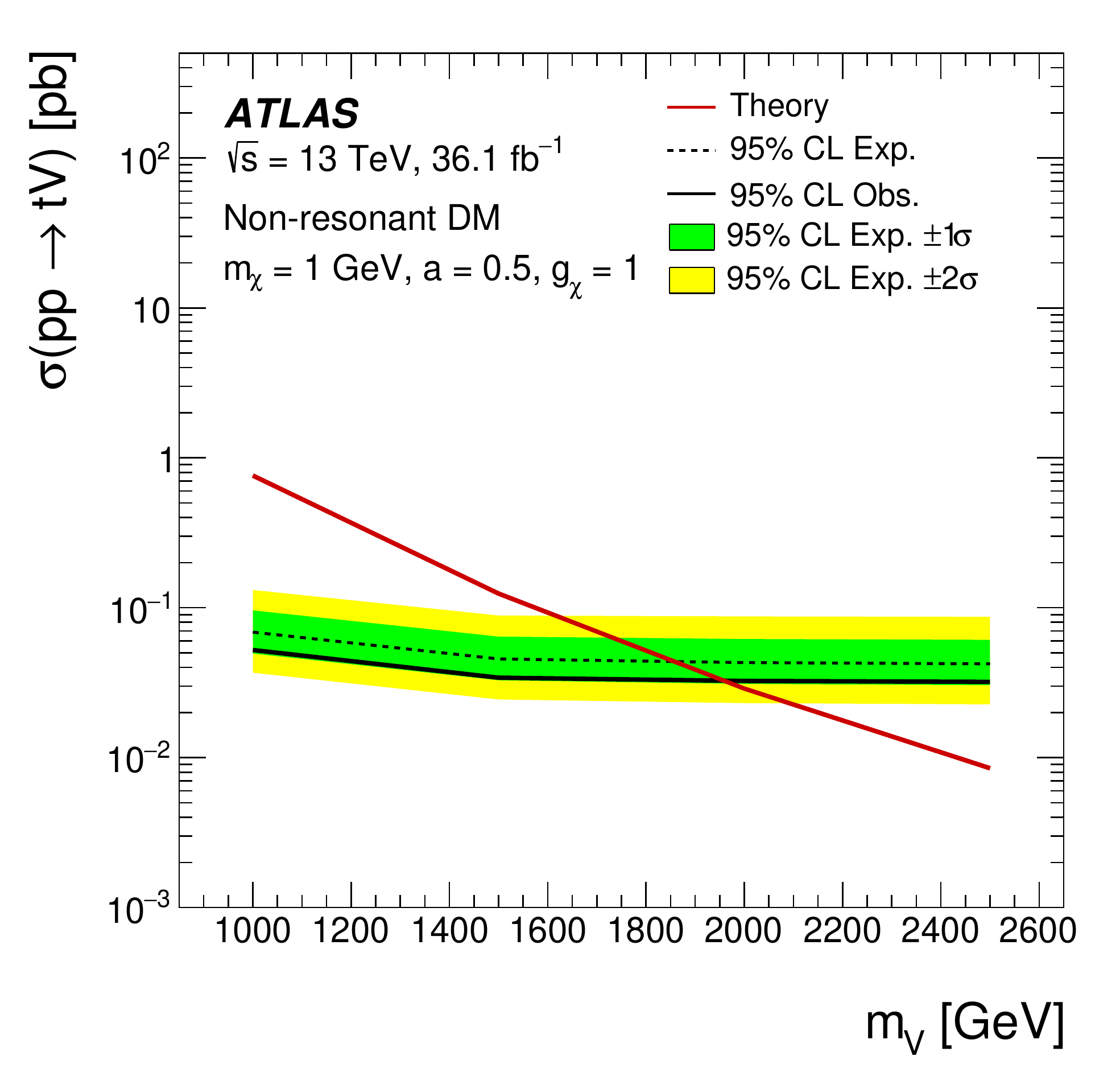} &
\includegraphics[width=.49\textwidth]{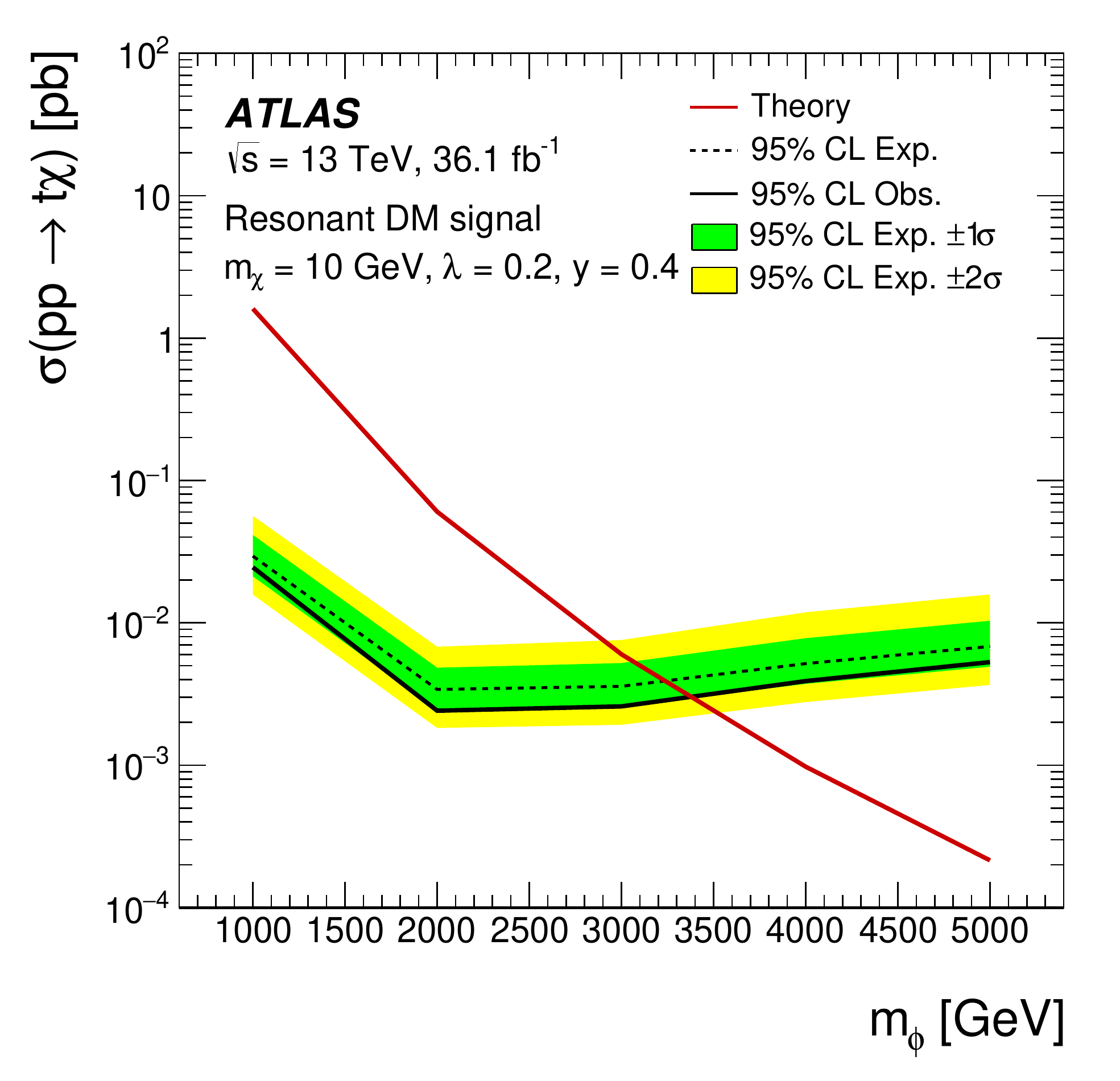}\\
(a) & (b) \\[5mm]
\multicolumn{2}{c}{\includegraphics[width=.49\textwidth]{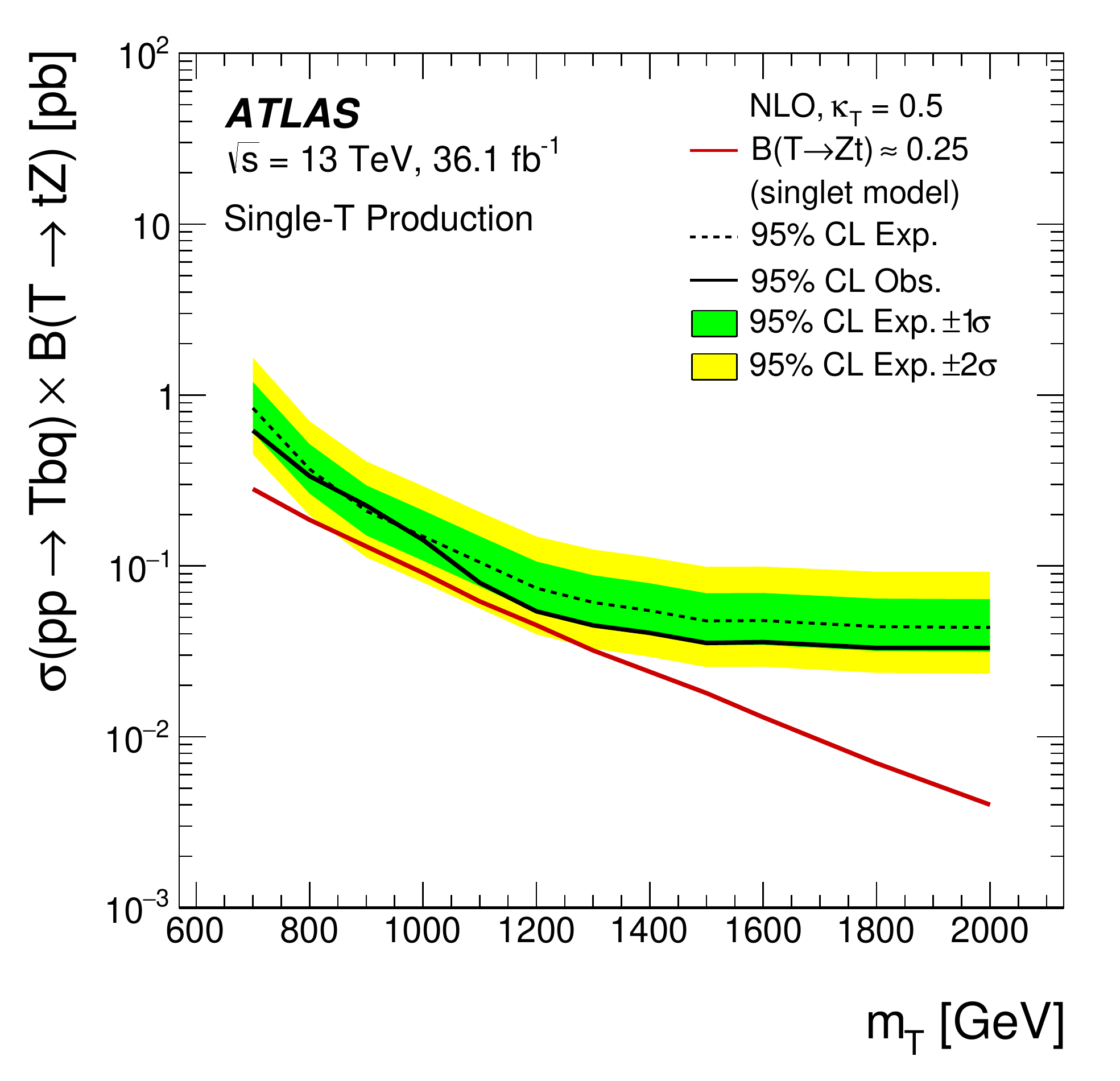}}\\
\multicolumn{2}{c}{(c)} \\
\end{tabular}
\end{center}
\caption{\label{fig:UpperLimitsCrossSection}
95\%{} CL upper limits on the signal cross-section as a function of (a) the $V$ mass in the non-resonant (NR) model, (b) the mass of the scalar
particle $\phi$ in the resonant (R) model and (c) the VLT mass. LO values for the production cross-section were computed for the non-resonant (resonant) DM production modes assuming $m_\chi=1$~\GeV, $a=0.5$ and $g_\chi=1$
($m_\chi=10$~\GeV, $\lambda = 0.2$ and $y=0.4$).
}
\end{figure}

Since there is no evidence of a signal, expected and observed upper
limits on the signal cross-section as a function of the $V$ mass for the non-resonant model, the mass of the scalar particle
$\phi$ for the resonant model and the VLT mass are derived at 95\%
confidence level (CL) and are shown in
\Fig{\ref{fig:UpperLimitsCrossSection}}. Comparing the cross-section
limits with the theoretical expectation, lower limits on the invisible
particle and VLT masses can be derived.
The LO values of the cross-section for non-resonant (resonant) DM
production are evaluated using \MGMCatNLO, as detailed in
Section~\ref{sec:data}, assuming $m_\chi=1$~\GeV, $a=0.5$ and
$g_\chi=1$ ($m_\chi=10$~\GeV, $\lambda = 0.2$ and $y=0.4$). For the VLT
interpretation, the single-$T$ production cross-section is taken from
the NLO calculations for $c_W = 1$, with the coupling $c_W$ defined in
Ref.~\cite{Matsedonskyi:2014mna}. The narrow-width approximation is
used~\cite{Matsedonskyi:2014mna}. For the current analysis, it was checked using dedicated Monte Carlo samples that the chirality of the coupling has negligible impact on the
considered observables. The cross-section is also corrected
for width effects calculated with \MGMCatNLO, assuming that the ratio
of NLO to LO cross-sections remains approximately the same for a
non-vanishing $T$-quark width. The computed cross-section is then
multiplied by the value of ${\mathcal{B}}(T\to Zt)$ in the singlet
model, which is $\approx 25$\% in the range of VLT masses investigated
in this analysis. The considered benchmark coupling of $\kappa_T =
0.5$ corresponds to $c_W=0.45$.
The observed (expected) mass limits at 95\%{} CL are 2.0 (1.9)~\TeV\ and
3.4 (3.3)~\TeV\ for the non-resonant and resonant dark-matter models,
respectively. For the VLT case, there is no observed or expected
mass exclusion for the considered reference benchmark coupling.

\begin{figure}[htbp]
\begin{center}
\begin{tabular}{c c}
\includegraphics[width=.45\textwidth]{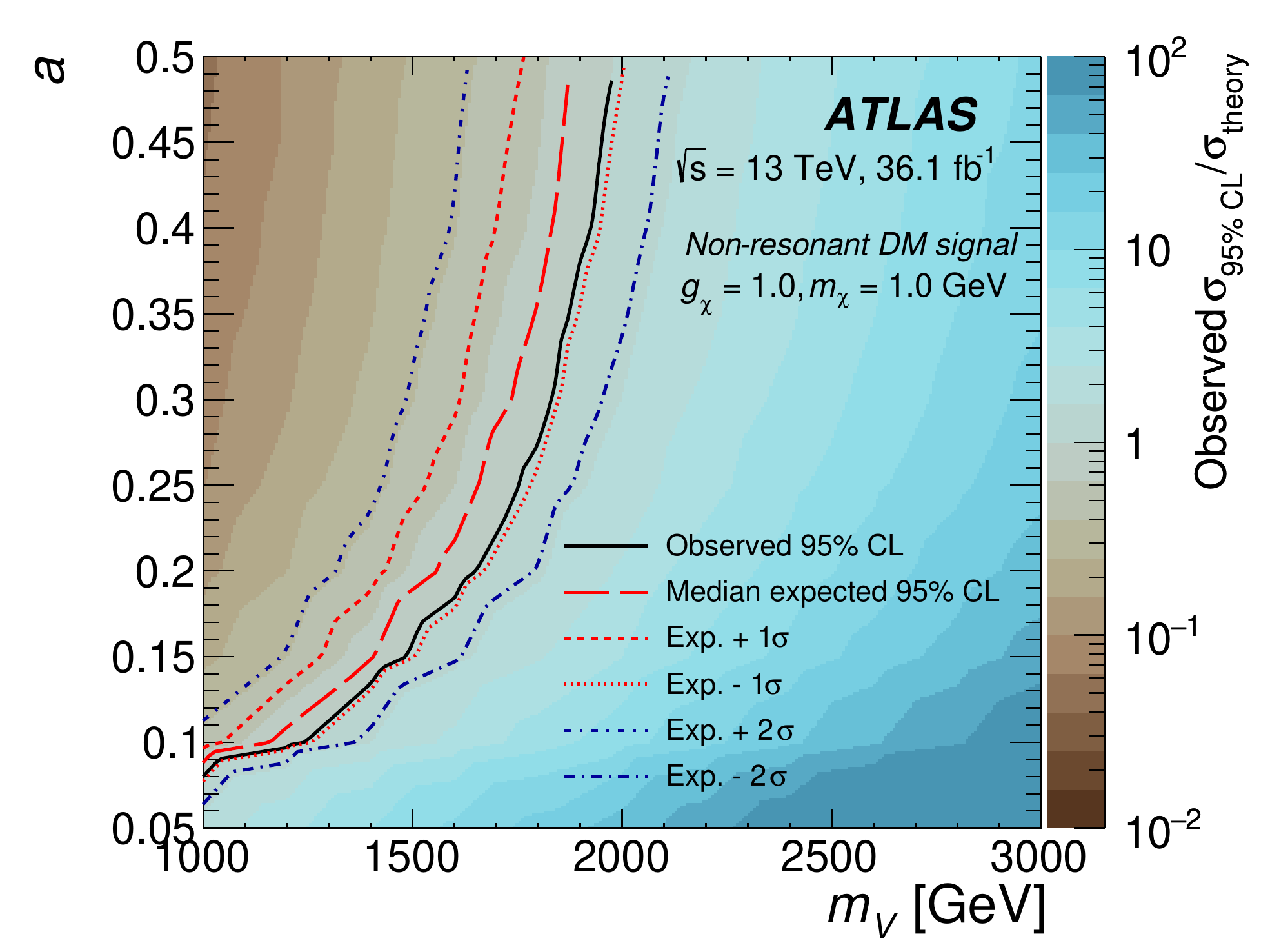} &
\includegraphics[width=.45\textwidth]{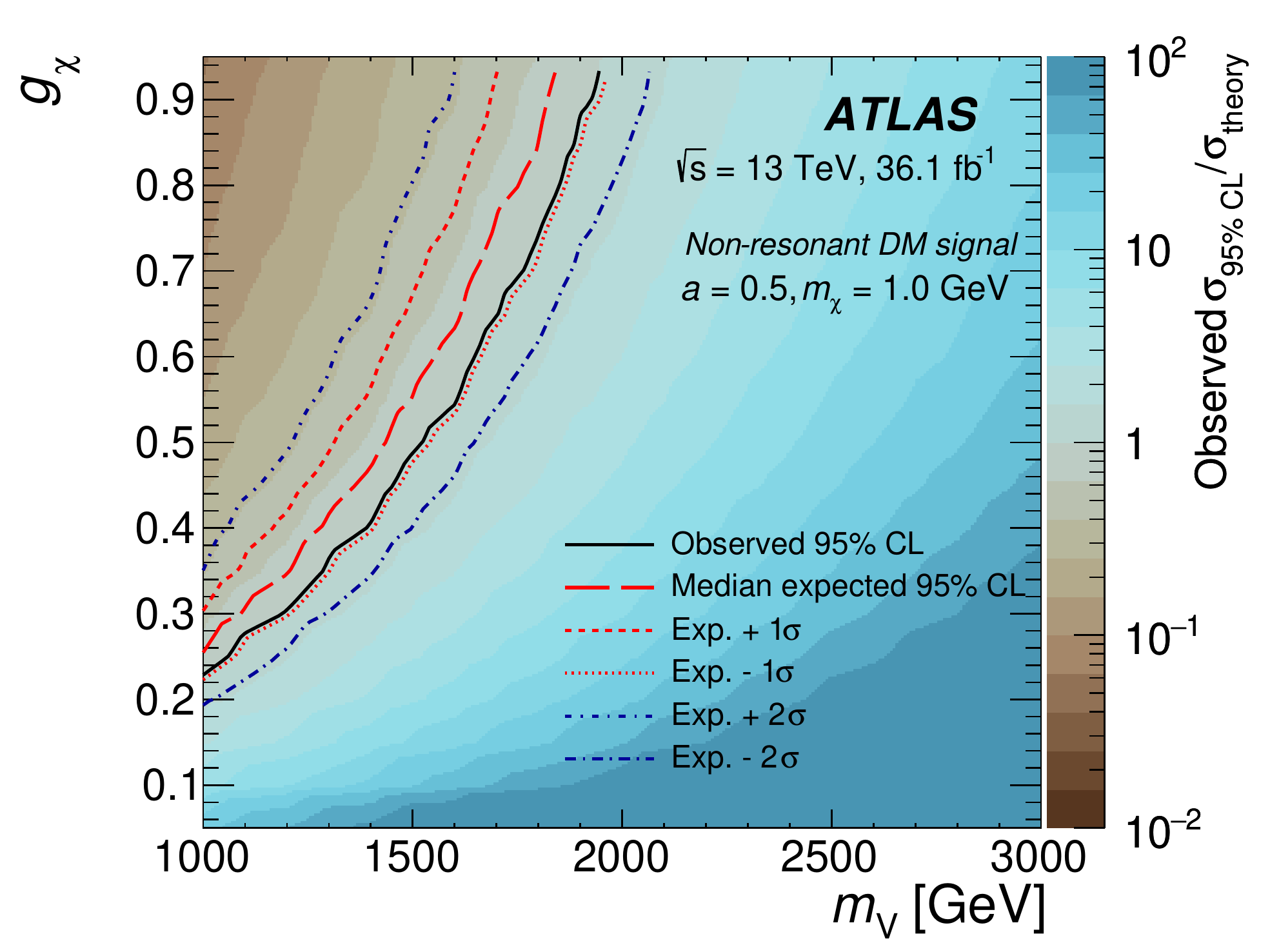} \\
(a) & (b) \\[5mm]
\multicolumn{2}{c}{\includegraphics[width=.45\textwidth]{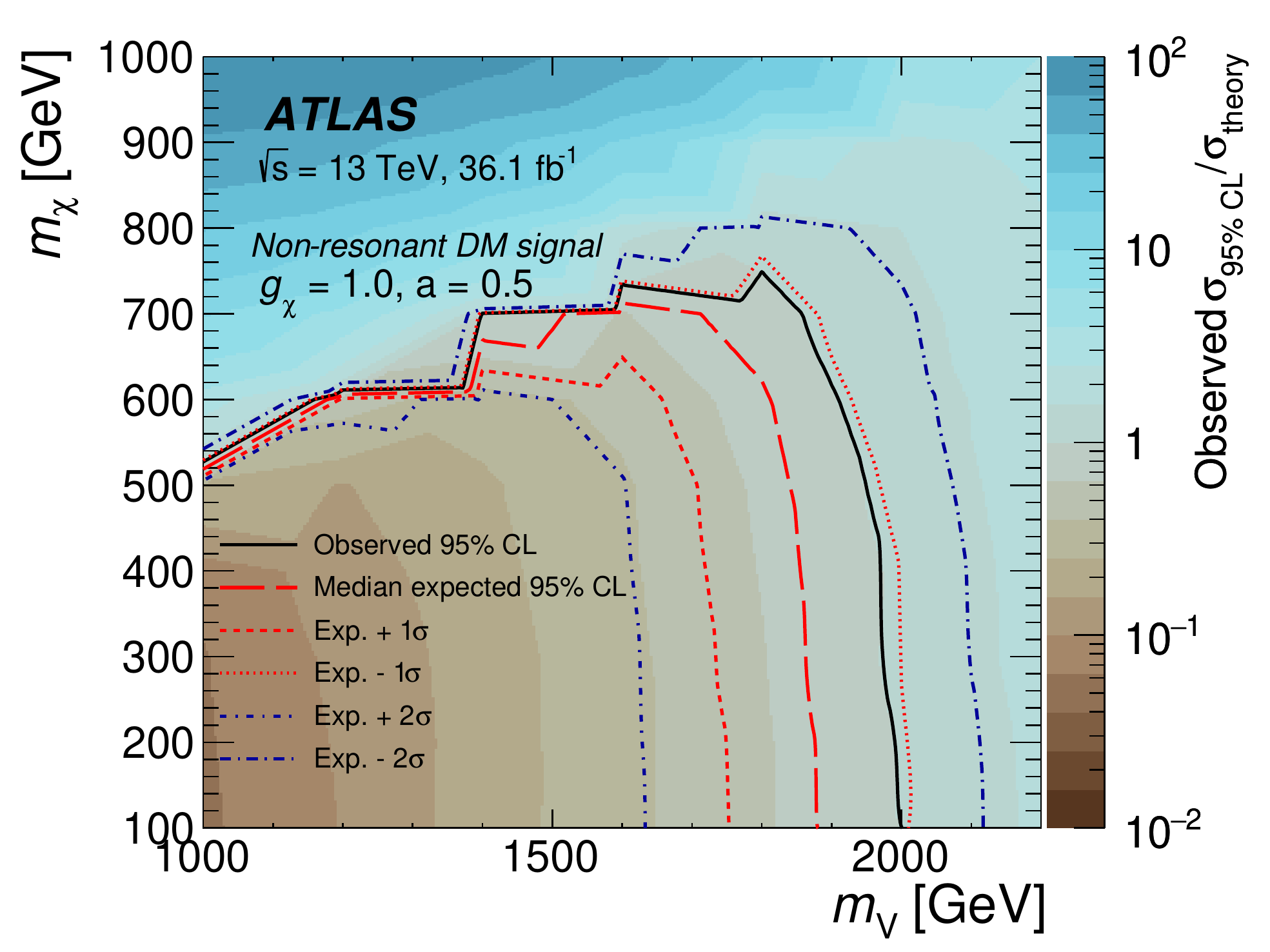}} \\
\multicolumn{2}{c}{(c)} \\[5mm]
\includegraphics[width=.45\textwidth]{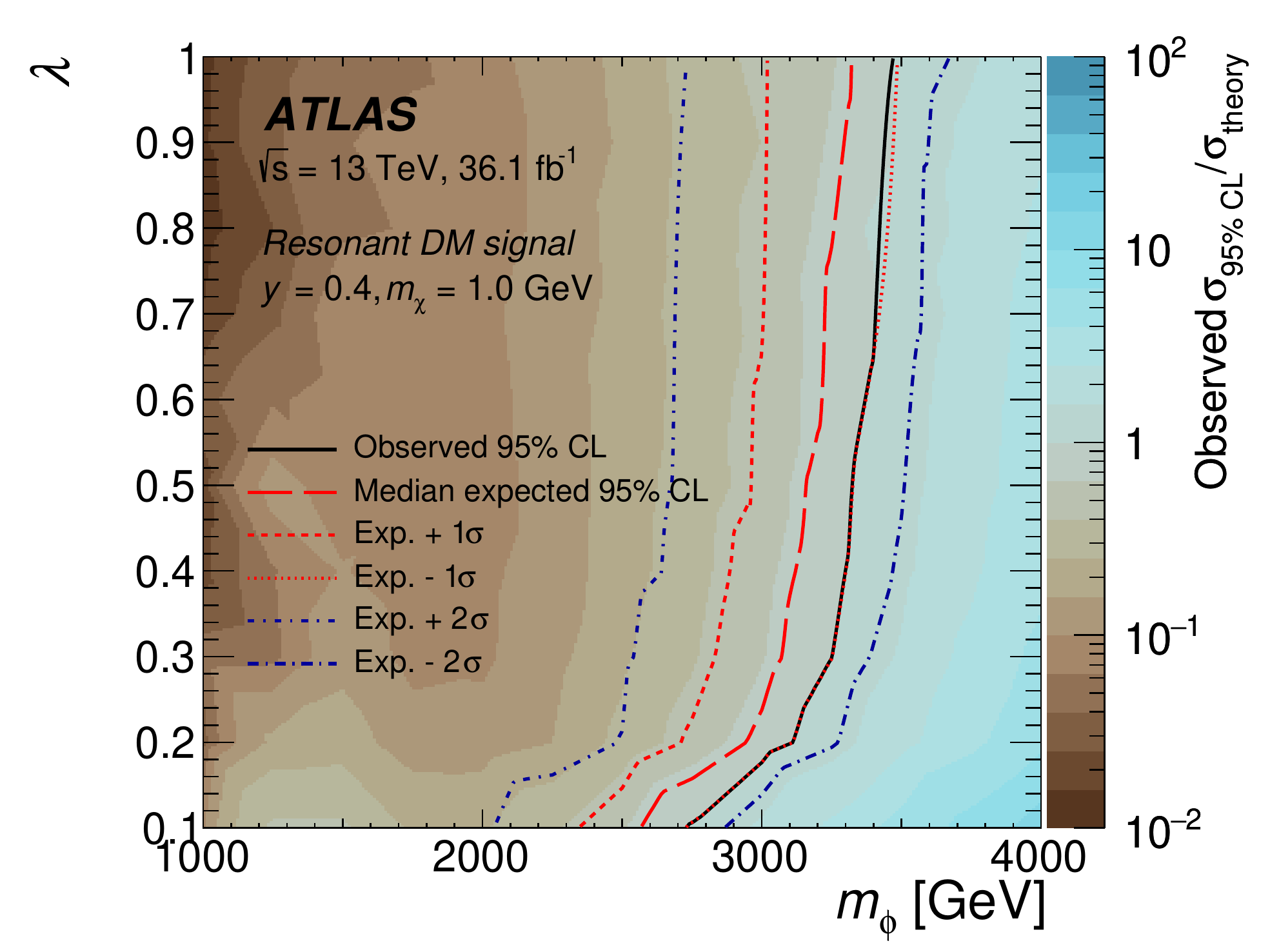} &
\includegraphics[width=.45\textwidth]{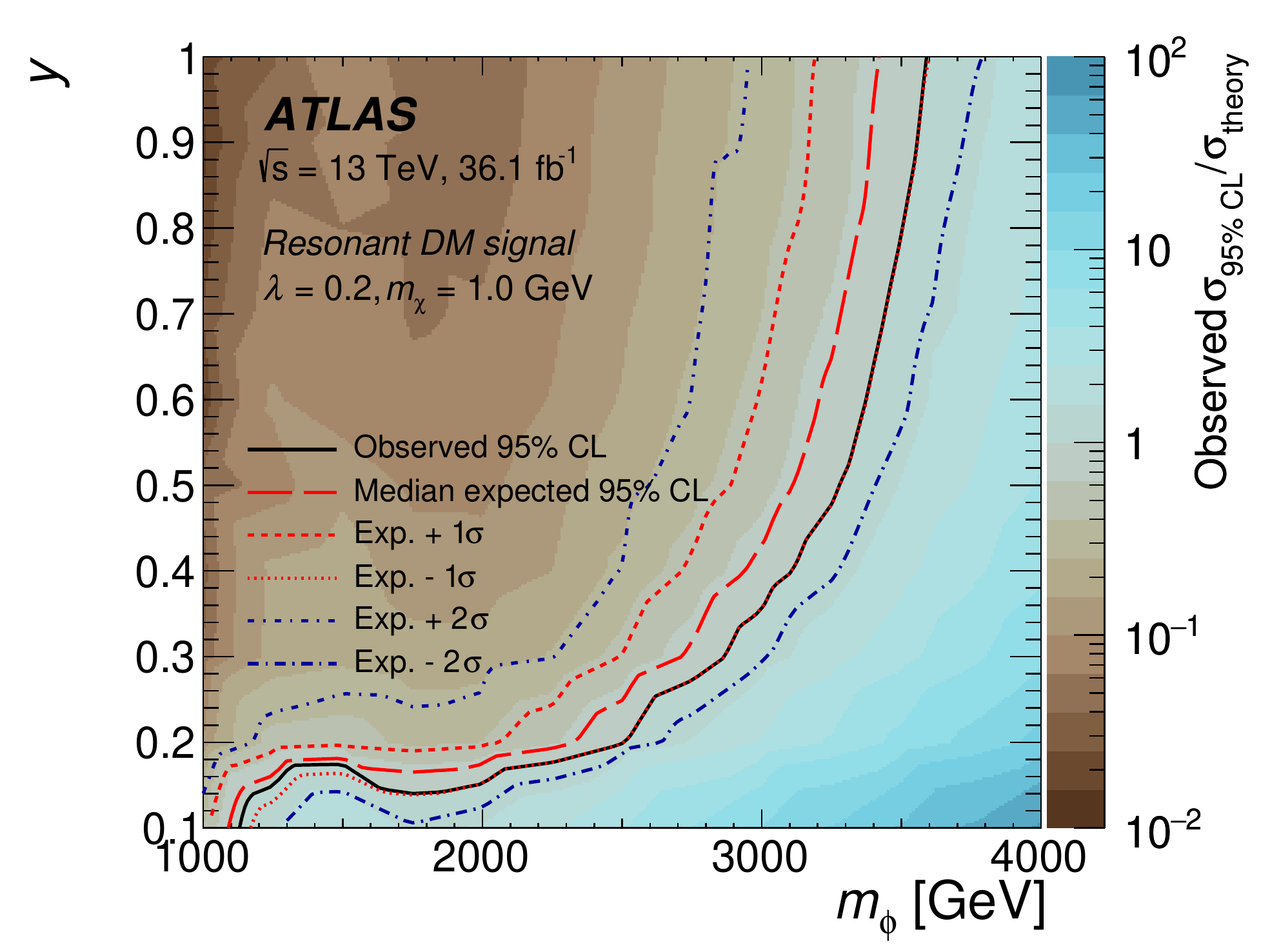} \\
(d) & (e) \\[5mm]
\end{tabular}
\end{center}
\caption{\label{fig:DMinterpretation}
The 95\% CL upper limit contours on the signal strength $\sigma/\sigma_\text{theory}$ are shown for
the non-resonant (NR) and resonant (R) DM production models. Non-resonant model: (a)
$V$ mass vs $a$; (b) $V$ mass vs $g_\chi$ and (c) $V$ mass vs mass of the DM candidate $\chi$.
Resonant model: (d) mass of the scalar $\phi$ vs $\lambda$; (e) mass of the scalar $\phi$ vs $y$.
The solid (dashed) lines
correspond to the observed (median expected and corresponding $\pm 1 \sigma$ and $\pm 2 \sigma$ bands) limits for $\sigma/\sigma_\text{theory}=1$.
The predicted cross-sections were computed with \MGMCatNLO.
}
\end{figure}

Two-dimensional exclusion
regions in the planes formed by the mediator masses, the DM particle mass, and
couplings between the DM, the new heavy particle and the SM fermions are obtained
by reweighting the events using the transverse momentum from the vector sum of the momenta of the DM candidates.
This procedure is validated with dedicated samples and
allows reproduction of the correct event kinematics for the masses and couplings required for
the multidimensional scans.
The observed (expected) 95\% CL upper limit contours for the signal
strength $\sigma/\sigma_\text{theory}$ are shown in
Figures~\ref{fig:DMinterpretation}(a)--\ref{fig:DMinterpretation}(c) for the non-resonant model,
in which $\sigma$ is the observed (expected) limit on the model
cross-section at a given point of the parameter space and
$\sigma_\text{theory}$ is the predicted cross-section in the model at
the same point. The corresponding results for the resonant model are shown in
Figures~\ref{fig:DMinterpretation}(d) and~\ref{fig:DMinterpretation}(e). Since a reweighting procedure was used
to obtain the required signal points, the results shown in Figure~\ref{fig:DMinterpretation}
include a systematic uncertainty in the signal normalisation associated with this procedure. This uncertainty
was estimated from dedicated MC samples to be 10\%{} and 25\%{} for the non-resonant and resonant case, respectively, by comparing reweighted samples with those generated with the corresponding signal masses and couplings.
 
\begin{figure}[htbp]
\begin{center}
\begin{tabular}{c}
\includegraphics[width=.57\textwidth]{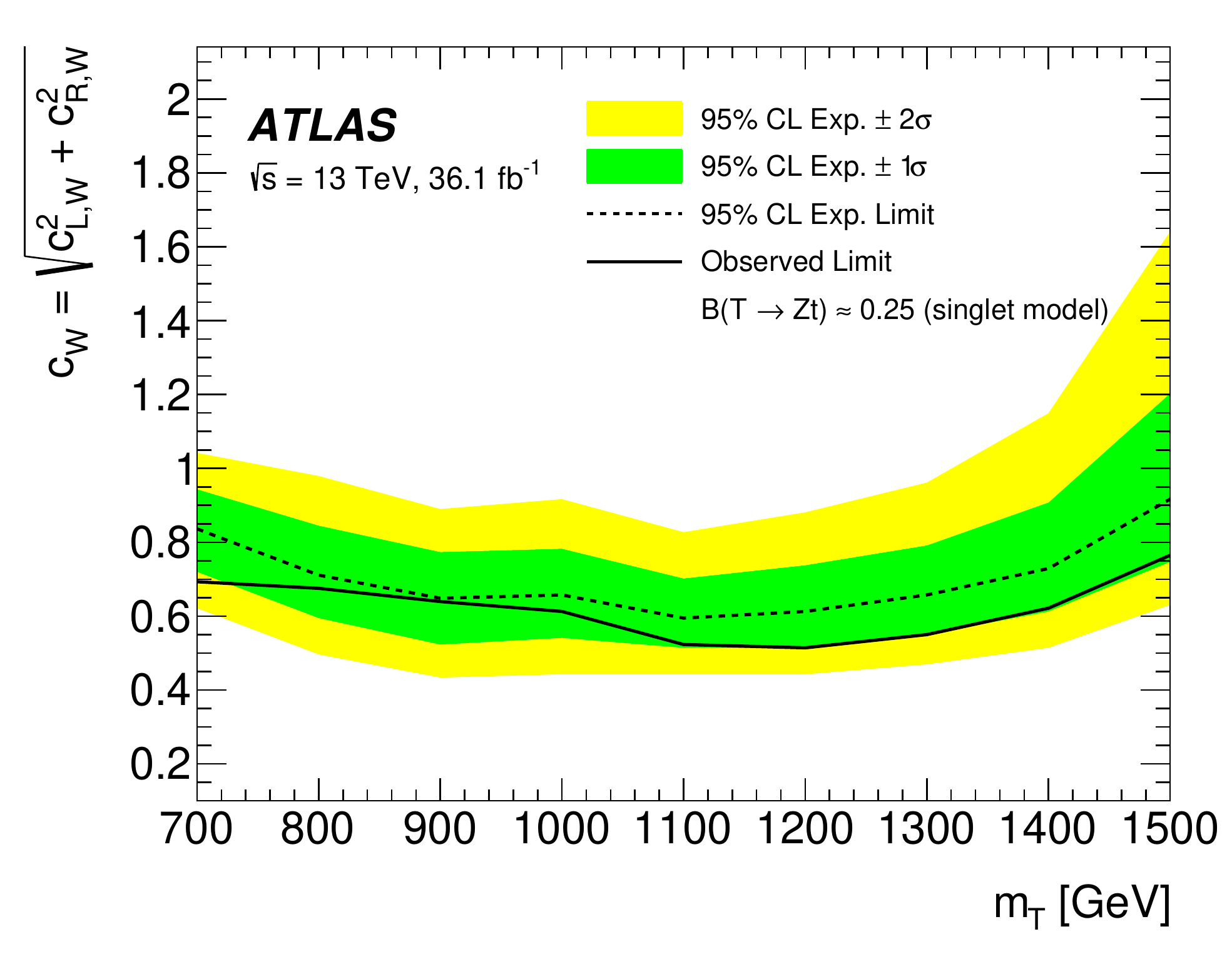} \\
(a) \\[5mm]
\includegraphics[width=.57\textwidth]{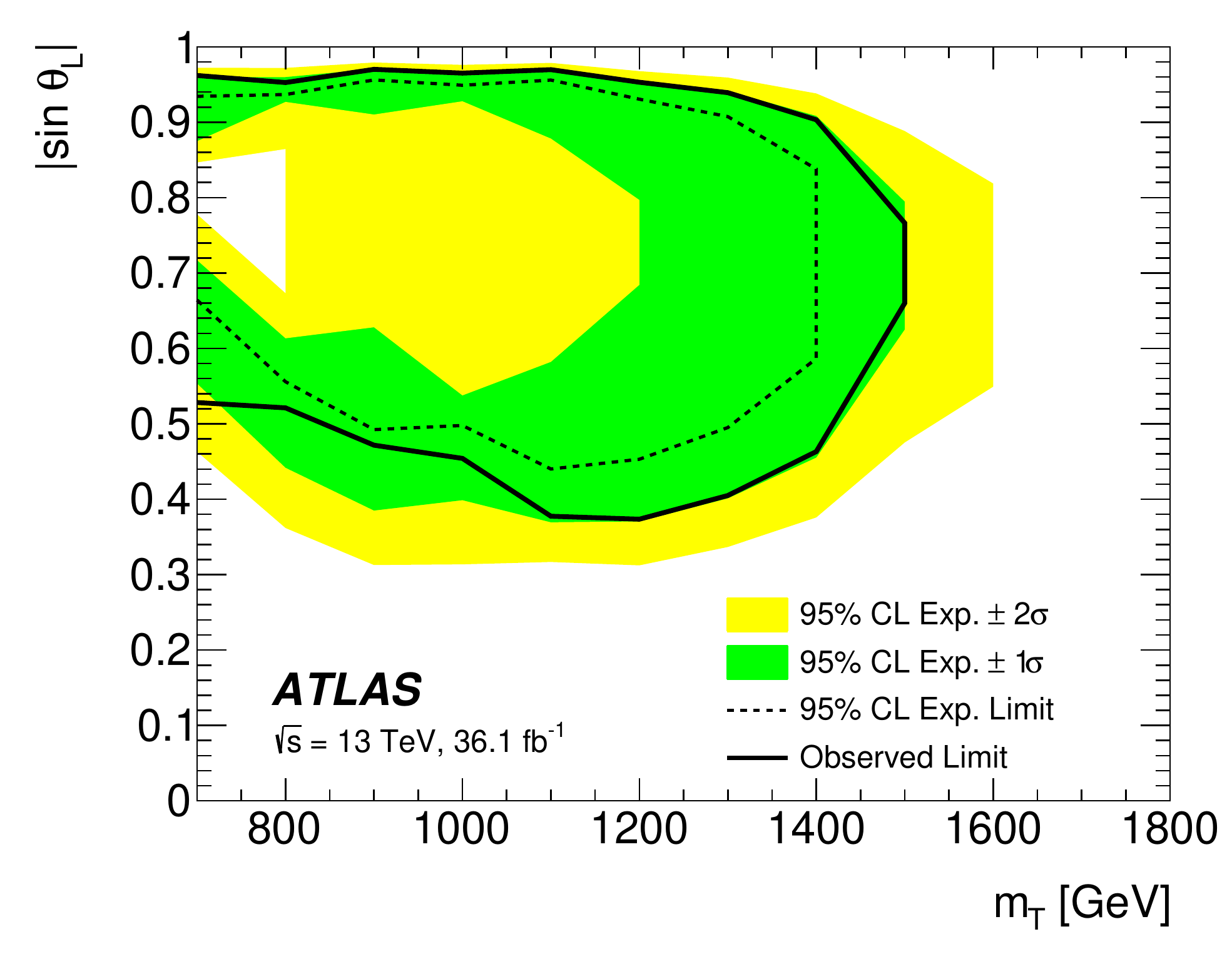} \\
(b) \\
\end{tabular}
\end{center}
\caption{\label{fig:VLTinterpretation}
Expected and observed 95\% CL limits from the combination of
the single-production channels on (a) the coupling of the $T$ quark to
SM particles, $c_W = \sqrt{c^2_{{\mathrm{L}},W} + c^2_{{\mathrm{R}},W}}$ assuming a singlet $T$, corresponding to a $\mathcal{B}$ of
$\approx 25$\%{}; and (b) the absolute value of $\sin\theta_{\mathrm{L}}$, with $\theta_{\mathrm{L}}$ being the mixing angle of a singlet $T$ with the SM top-quark.
}

\end{figure}

The limited sensitivity of the current analysis to single VLT production for low $T$ masses (cf.\
\Fig{\ref{fig:UpperLimitsCrossSection}(c)}) implies that there is also less
sensitivity to the corresponding coupling. This can be seen in
\Fig{\ref{fig:VLTinterpretation}(a)}, which shows the expected and observed 95\%{}
CL upper limits on $c_W$, taken as the sum in quadrature of the
left- and right-handed couplings $c_{{\mathrm{L}},W}$ and $c_{{\mathrm{R}},W}$, as a function of the VLT
mass. Nonetheless, the sensitivity remains approximately constant for masses up to
1.4~\TeV. A singlet $T$, which corresponds to ${\mathcal{B}}(T\to Zt) \approx 25$\% over
the mass range studied in this analysis, was assumed. The obtained limits on $c_W$
can also be translated into expected and observed 95\% CL upper limits for the
mixing angle of a singlet $T$ with the SM top-quark, as shown in
\Fig{\ref{fig:VLTinterpretation}(b)}. For these results, a signal reweighting
procedure was adopted in order to take into account the width effects induced by
the variation of the $c_W$ coupling. The systematic uncertainty in the signal
normalisation was estimated to be 3\%{} from dedicated MC samples and was
considered when deriving the limits shown in \Fig{\ref{fig:VLTinterpretation}}. In
the range $m(T)>1.1$~\TeV, the obtained exclusion limit on the $c_W$ coupling improves
on the previous results~\cite{Aaboud:2018saj}.

\clearpage
 
\section{Conclusions}
\label{sec:conclusions}
 
This analysis seeks anomalous production of events with large \met and a single top-quark in LHC $pp$ data at $\sqrt{s}=13$~\TeV\ collected by the ATLAS detector in 2015 and 2016, corresponding to an integrated luminosity of 36.1~fb$^{-1}$. No deviations with respect to SM predictions are observed and 95\%{} CL upper limits on the production cross-section of three BSM processes are obtained: resonant and non-resonant production of dark matter (DM) in association with single top-quarks, and single production of vector-like $T$ quarks decaying into $tZ(\to\nu\bar\nu)$.
 
These limits are also interpreted in terms of the excluded regions in the parameter space of the considered BSM scenarios. For DM production in the non-resonant scenario, masses of a new vector particle coupling to the DM candidate up to 2~\TeV\ are excluded at 95\%{} CL for  $m_\chi=1$~\GeV, $g_\chi=1.0$ and $a=0.5$, while in the resonant case, masses of a new scalar particle coupling to DM up to 3.4~\TeV\ are excluded at 95\%{} CL for $m_\chi=10$~\GeV, $y=0.4$ and $\lambda=0.2$. For the production of $T$ singlets, couplings of these new quarks to top-quarks and $W$ bosons, $c_W$, above 0.7 are excluded for $m_T=1.4$~\TeV\ and below.
 
\appendix
\section{Event yields in the signal and control regions after the fit to data}

The numbers of events observed in the signal and control regions are
presented in
Tables~\ref{tab:PostFitYieldsNRDM}, \ref{tab:PostFitYieldsRDM} and
\ref{tab:PostFitYieldsVLT}, together with the backgrounds estimated in
the simultaneous fit to data in the corresponding regions under the
background-only hypothesis. In Table~\ref{tab:PostFitYieldsNRDM}, 1L and 0L DM signal regions as well as the 1L and 0L control regions are included in the fit. Table~\ref{tab:PostFitYieldsRDM} includes 0L DM signal and control regions, while VLT signal and control regions are considered for Table~\ref{tab:PostFitYieldsVLT}.
 
\begin{table}[htbp]
\begin{center}
\begin{tabular}{l |r@{~$\pm$~}l  r@{~$\pm$~}l  r@{~$\pm$~}l}
\hline \hline
Leptonic channel  & \multicolumn{2}{c}{1L-DM-SR} & \multicolumn{2}{c}{1L-TCR} & \multicolumn{2}{c}{1L-WCR} \\
\hline
$t\bar{t}$   & \num[round-mode=figures,round-precision=3]{385.269} & \num[round-mode=figures,round-precision=2]{40.8536} & \num[round-mode=figures,round-precision=3]{12130.6} & \num[round-mode=figures,round-precision=2]{1960.85} & \num[round-mode=figures,round-precision=3]{8
467.17} & \num[round-mode=figures,round-precision=2]{798.86}  \\
 
Non-$t\bar{t}$   & \num[round-mode=figures,round-precision=3]{116.639} & \num[round-mode=figures,round-precision=2]{17.4398} & \num[round-mode=figures,round-precision=3]{5543.64} & \num[round-mode=figures,round-precision=2]{963.185} & \num[round-mode=figures,round-precision=3]{118664} & \num[round-mode=figures,round-precision=2]{25865.4} \\
 
\hline
Total  & \num[round-mode=figures,round-precision=3]{501.908} & \num[round-mode=figures,round-precision=2]{62.3669} & \num[round-mode=figures,round-precision=3]{17674.2} & \num[round-mode=figures,round-precision=2]{3055.84} & \num[round-mode=figures,round-precision=3]{127131} & \num[round-mode=figures,round-precision=2]{25826.7} \\
\hline
Data   & \multicolumn{2}{c}{511} & \multicolumn{2}{c}{17662} & \multicolumn{2}{c}{127286}  \\ \hline
\hline \hline
Hadronic channel & \multicolumn{2}{c}{0L-DM-SR} & \multicolumn{2}{c}{0L-TCR} & \multicolumn{2}{c}{0L-VCR}\\
\hline
$t\bar{t}$   & \num[round-mode=figures,round-precision=3]{9899.73} & \num[round-mode=figures,round-precision=2]{874.877} & \num[round-mode=figures,round-precision=3]{7155.79} & \num[round-mode=figures,round-precision=2]{618.559} & \num[round-mode=figures,round-precision=3]{5901.09} & \num[round-mode=figures,round-precision=2]{251.421} \\
 
Single top   &\num[round-mode=figures,round-precision=2]{990.627} & \num[round-mode=figures,round-precision=2]{105.467} & \num[round-mode=figures,round-precision=3]{273.158} & \num[round-mode=figures,round-precision=2]{35.6454} & \num[round-mode=figures,round-precision=3]{879.088} & \num[round-mode=figures,round-precision=2]{98.3525} \\
 
$W$+jets   &\num[round-mode=figures,round-precision=3]{2047.73} & \num[round-mode=figures,round-precision=2]{519.188} & \num[round-mode=figures,round-precision=3]{118.843} & \num[round-mode=figures,round-precision=2]{65.3012} & \num[round-mode=figures,round-precision=3]{23075.5} & \num[round-mode=figures,round-precision=2]{4919.78} \\
 
$Z$+jets   &\num[round-mode=figures,round-precision=3]{2455.82} & \num[round-mode=figures,round-precision=2]{461.385} & \num[round-mode=figures,round-precision=3]{135.397} & \num[round-mode=figures,round-precision=2]{60.6836} & \num[round-mode=figures,round-precision=3]{29915.4} & \num[round-mode=figures,round-precision=2]{4607.27} \\
 
Multijet   &\num[round-mode=figures,round-precision=2]{86.7035} & \num[round-mode=figures,round-precision=2]{90.1524} & \num[round-mode=figures,round-precision=2]{757.292} & \num[round-mode=figures,round-precision=2]{348.217} & \num[round-mode=figures,round-precision=2]{0} & \num[round-mode=figures,round-precision=2]{0} \\
 
Other   &\num[round-mode=figures,round-precision=3]{328.204} & \num[round-mode=figures,round-precision=2]{40.8788} & \num[round-mode=figures,round-precision=3]{50.0684} & \num[round-mode=figures,round-precision=2]{5.55382} & \num[round-mode=figures,round-precision=3]{2667.09} & \num[round-mode=figures,round-precision=2]{305.944} \\
\hline
Total  &  \num[round-mode=figures,round-precision=3]{15808.8} & \num[round-mode=figures,round-precision=2]{1170.09} & \num[round-mode=figures,round-precision=3]{8490.54} & \num[round-mode=figures,round-precision=2]{761.173} & \num[round-mode=figures,round-precision=3]{62438.1} & \num[round-mode=figures,round-precision=2]{1485.9} \\
\hline
Data   & \multicolumn{2}{c}{15781} &  \multicolumn{2}{c}{8493} & \multicolumn{2}{c}{62304} \\
\hline \hline
\end{tabular} 
\caption{Numbers of events observed in the signal and control regions used for the non-resonant dark-matter search,
together with the estimated SM backgrounds in the fit to data, under the background-only
hypothesis.
The uncertainties include statistical and
systematic uncertainties. The uncertainties in the individual
backgrounds are correlated, and do not necessarily add in quadrature
to the total background uncertainty.
}
\label{tab:PostFitYieldsNRDM}
\end{center}
\end{table}
 
\begin{table}[htbp]
\begin{center}
\begin{tabular}{l |r@{~$\pm$~}l  r@{~$\pm$~}l  r@{~$\pm$~}l}
\hline\hline
& \multicolumn{2}{c}{0L-DM-SR} & \multicolumn{2}{c}{0L-TCR} & \multicolumn{2}{c}{0L-VCR}\\
\hline
$t\bar{t}$   & \num[round-mode=figures,round-precision=3]{9689.94} & \num[round-mode=figures,round-precision=2]{615.037} & \num[round-mode=figures,round-precision=3]{7109.08} & \num[round-mode=figures,round-precision=2]{460.894} & \num[round-mode=figures,round-precision=3]{5711.87} & \num[round-mode=figures,round-precision=2]{583.489} \\
Single top   & \num[round-mode=figures,round-precision=2]{986.664} & \num[round-mode=figures,round-precision=2]{109.637} & \num[round-mode=figures,round-precision=3]{281.51} & \num[round-mode=figures,round-precision=1]{38.5556}  & \num[round-mode=figures,round-precision=2]{869.132} & \num[round-mode=figures,round-precision=2]{107.291} \\
$W$+jets       & \num[round-mode=figures,round-precision=3]{2066.4} & \num[round-mode=figures,round-precision=2]{539.021}  & \num[round-mode=figures,round-precision=3]{120.77} & \num[round-mode=figures,round-precision=2]{67.1422}  & \num[round-mode=figures,round-precision=3]{22993.4} & \num[round-mode=figures,round-precision=2]{5015.16} \\
$Z$+jets       & \num[round-mode=figures,round-precision=3]{2609.67} & \num[round-mode=figures,round-precision=2]{526.734} & \num[round-mode=figures,round-precision=3]{149.047} & \num[round-mode=figures,round-precision=2]{61.0072} & \num[round-mode=figures,round-precision=3]{30058.7} & \num[round-mode=figures,round-precision=2]{4653.8}  \\
Other        & \num[round-mode=figures,round-precision=3]{330.293} & \num[round-mode=figures,round-precision=2]{44.1154} & \num[round-mode=figures,round-precision=3]{51.4163} & \num[round-mode=figures,round-precision=2]{6.11536} & \num[round-mode=figures,round-precision=3]{2674.89} & \num[round-mode=figures,round-precision=2]{314.522} \\
Multijet     & \num[round-mode=figures,round-precision=2]{92.1532} & \num[round-mode=figures,round-precision=2]{88.2063} & \num[round-mode=figures,round-precision=2]{796.115} & \num[round-mode=figures,round-precision=2]{359.483} & \num[round-mode=figures,round-precision=2]{0} & \num[round-mode=figures,round-precision=2]{0} \\
\hline
Total        & \num[round-mode=figures,round-precision=3]{15775.1} & \num[round-mode=figures,round-precision=2]{365.001} & \num[round-mode=figures,round-precision=3]{8507.94} & \num[round-mode=figures,round-precision=2]{277.792} & \num[round-mode=figures,round-precision=3]{62307.9} & \num[round-mode=figures,round-precision=2]{1416.3}  \\
\hline
Data   & \multicolumn{2}{c}{15781} & \multicolumn{2}{c}{8493} & \multicolumn{2}{c}{62304} \\
\hline \hline
\end{tabular}
\caption{Numbers of events observed in the signal and control regions used for the resonant dark-matter search,
together with the estimated SM backgrounds in the fit to data, under the background-only
hypothesis. The uncertainties include statistical and
systematic uncertainties. The uncertainties in the individual
backgrounds are correlated, and do not necessarily add in quadrature
to the total background uncertainty.}
\label{tab:PostFitYieldsRDM}
\end{center}
\end{table}

\begin{table}[htbp]
\begin{center}
\begin{tabular}{l |r@{~$\pm$~}l  r@{~$\pm$~}l  r@{~$\pm$~}l}
\hline\hline
& \multicolumn{2}{c}{0L-VLT-SR} & \multicolumn{2}{c}{0L-TCR} & \multicolumn{2}{c}{0L-VCR}\\
\hline
$t\bar{t}$   & \num[round-mode=figures,round-precision=3]{3556.56} & \num[round-mode=figures,round-precision=2]{277.848} & \num[round-mode=figures,round-precision=3]{7160.43} & \num[round-mode=figures,round-precision=2]{374.516}   & \num[round-mode=figures,round-precision=3]{5312.44} & \num[round-mode=figures,round-precision=2]{739.719} \\
Single top   & \num[round-mode=figures,round-precision=3]{322.749} & \num[round-mode=figures,round-precision=2]{44.9592} & \num[round-mode=figures,round-precision=3]{278.286} & \num[round-mode=figures,round-precision=2]{36.603}    & \num[round-mode=figures,round-precision=2]{818.348} & \num[round-mode=figures,round-precision=2]{123.198} \\
$W$+jets       & \num[round-mode=figures,round-precision=2]{664.53} & \num[round-mode=figures,round-precision=2]{198.605}  & \num[round-mode=figures,round-precision=3]{125.799} & \num[round-mode=figures,round-precision=2]{71.8749}   & \num[round-mode=figures,round-precision=3]{21856.4} & \num[round-mode=figures,round-precision=2]{5871.73} \\
$Z$+jets       & \num[round-mode=figures,round-precision=2]{832.087} & \num[round-mode=figures,round-precision=2]{182.492} & \num[round-mode=figures,round-precision=3]{159.518} & \num[round-mode=figures,round-precision=2]{63.8904}   & \num[round-mode=figures,round-precision=3]{31770} & \num[round-mode=figures,round-precision=2]{4769.39}   \\
Other        & \num[round-mode=figures,round-precision=2]{82.0945} & \num[round-mode=figures,round-precision=2]{13.8548} &
\num[round-mode=figures,round-precision=2]{801.2797} & \num[round-mode=figures,round-precision=2]{316.9771} &
\num[round-mode=figures,round-precision=3]{2590.03} & \num[round-mode=figures,round-precision=2]{341.838} \\
\hline
Total        & \num[round-mode=figures,round-precision=3]{5458.02} & \num[round-mode=figures,round-precision=2]{161.483} & \num[round-mode=figures,round-precision=3]{8525.31} & \num[round-mode=figures,round-precision=2]{273.325}   & \num[round-mode=figures,round-precision=3]{62347.2} & \num[round-mode=figures,round-precision=2]{1440.4}  \\
\hline
Data   & \multicolumn{2}{c}{5454} & \multicolumn{2}{c}{8493} & \multicolumn{2}{c}{62304} \\
\hline \hline
\end{tabular}
\caption{Numbers of events observed in the signal and control regions used for the vector-like $T$-quark search,
together with the estimated SM backgrounds in the
fit to data, under the background-only
hypothesis. The uncertainties include statistical and
systematic uncertainties. The uncertainties in the individual
backgrounds are correlated, and do not necessarily add in quadrature
to the total background uncertainty.}
\label{tab:PostFitYieldsVLT}
\end{center}
\end{table}

\section*{Acknowledgements}
 
 
We thank CERN for the very successful operation of the LHC, as well as the
support staff from our institutions without whom ATLAS could not be
operated efficiently.
 
We acknowledge the support of ANPCyT, Argentina; YerPhI, Armenia; ARC, Australia; BMWFW and FWF, Austria; ANAS, Azerbaijan; SSTC, Belarus; CNPq and FAPESP, Brazil; NSERC, NRC and CFI, Canada; CERN; CONICYT, Chile; CAS, MOST and NSFC, China; COLCIENCIAS, Colombia; MSMT CR, MPO CR and VSC CR, Czech Republic; DNRF and DNSRC, Denmark; IN2P3-CNRS, CEA-DRF/IRFU, France; SRNSFG, Georgia; BMBF, HGF, and MPG, Germany; GSRT, Greece; RGC, Hong Kong SAR, China; ISF and Benoziyo Center, Israel; INFN, Italy; MEXT and JSPS, Japan; CNRST, Morocco; NWO, Netherlands; RCN, Norway; MNiSW and NCN, Poland; FCT, Portugal; MNE/IFA, Romania; MES of Russia and NRC KI, Russian Federation; JINR; MESTD, Serbia; MSSR, Slovakia; ARRS and MIZ\v{S}, Slovenia; DST/NRF, South Africa; MINECO, Spain; SRC and Wallenberg Foundation, Sweden; SERI, SNSF and Cantons of Bern and Geneva, Switzerland; MOST, Taiwan; TAEK, Turkey; STFC, United Kingdom; DOE and NSF, United States of America. In addition, individual groups and members have received support from BCKDF, CANARIE, CRC and Compute Canada, Canada; COST, ERC, ERDF, Horizon 2020, and Marie Sk{\l}odowska-Curie Actions, European Union; Investissements d' Avenir Labex and Idex, ANR, France; DFG and AvH Foundation, Germany; Herakleitos, Thales and Aristeia programmes co-financed by EU-ESF and the Greek NSRF, Greece; BSF-NSF and GIF, Israel; CERCA Programme Generalitat de Catalunya, Spain; The Royal Society and Leverhulme Trust, United Kingdom.
 
The crucial computing support from all WLCG partners is acknowledged gratefully, in particular from CERN, the ATLAS Tier-1 facilities at TRIUMF (Canada), NDGF (Denmark, Norway, Sweden), CC-IN2P3 (France), KIT/GridKA (Germany), INFN-CNAF (Italy), NL-T1 (Netherlands), PIC (Spain), ASGC (Taiwan), RAL (UK) and BNL (USA), the Tier-2 facilities worldwide and large non-WLCG resource providers. Major contributors of computing resources are listed in Ref.~\cite{ATL-GEN-PUB-2016-002}.
 

\clearpage
\printbibliography

\clearpage
 
\begin{flushleft}
{\Large The ATLAS Collaboration}

\bigskip

M.~Aaboud$^\textrm{\scriptsize 34d}$,    
G.~Aad$^\textrm{\scriptsize 99}$,    
B.~Abbott$^\textrm{\scriptsize 125}$,    
O.~Abdinov$^\textrm{\scriptsize 13,*}$,    
B.~Abeloos$^\textrm{\scriptsize 129}$,    
D.K.~Abhayasinghe$^\textrm{\scriptsize 91}$,    
S.H.~Abidi$^\textrm{\scriptsize 164}$,    
O.S.~AbouZeid$^\textrm{\scriptsize 39}$,    
N.L.~Abraham$^\textrm{\scriptsize 153}$,    
H.~Abramowicz$^\textrm{\scriptsize 158}$
% ...
\bigskip
\\

% ...
$^{13}$Institute of Physics, Azerbaijan Academy of Sciences, Baku; Azerbaijan.\\
% ...
$^{34}$$^{(a)}$Facult\'e des Sciences Ain Chock, R\'eseau Universitaire de Physique des Hautes Energies - Universit\'e Hassan II, Casablanca;$^{(b)}$Centre National de l'Energie des Sciences Techniques Nucleaires (CNESTEN), Rabat;$^{(c)}$Facult\'e des Sciences Semlalia, Universit\'e Cadi Ayyad, LPHEA-Marrakech;$^{(d)}$Facult\'e des Sciences, Universit\'e Mohamed Premier and LPTPM, Oujda;$^{(e)}$Facult\'e des sciences, Universit\'e Mohammed V, Rabat; Morocco.\\
% ...
$^{39}$Niels Bohr Institute, University of Copenhagen, Copenhagen; Denmark.\\
% ...
$^{91}$Department of Physics, Royal Holloway University of London, Egham; United Kingdom.\\
% ...
$^{99}$CPPM, Aix-Marseille Universit\'e, CNRS/IN2P3, Marseille; France.\\
% ...
$^{125}$Homer L. Dodge Department of Physics and Astronomy, University of Oklahoma, Norman OK; United States of America.\\
% ...
$^{129}$LAL, Universit\'e Paris-Sud, CNRS/IN2P3, Universit\'e Paris-Saclay, Orsay; France.\\
% ...
$^{153}$Department of Physics and Astronomy, University of Sussex, Brighton; United Kingdom.\\
% ...
$^{158}$Raymond and Beverly Sackler School of Physics and Astronomy, Tel Aviv University, Tel Aviv; Israel.\\
% ...
$^{164}$Department of Physics, University of Toronto, Toronto ON; Canada.\\
% ...
$^{a}$ Also at Borough of Manhattan Community College, City University of New York, NY; United States of America.\\
$^{b}$ Also at California State University, East Bay; United States of America.\\
$^{c}$ Also at Centre for High Performance Computing, CSIR Campus, Rosebank, Cape Town; South Africa.\\
$^{d}$ Also at CERN, Geneva; Switzerland.\\
$^{e}$ Also at CPPM, Aix-Marseille Universit\'e, CNRS/IN2P3, Marseille; France.\\
% ...
$^{*}$ Deceased

\end{flushleft}


\clearpage

\end{document}